%% file: main-short.tex
\documentclass[twocolumn]{aastex63}
\usepackage[utf8]{inputenc}
\usepackage{graphicx}
\usepackage{color}
\usepackage{amsmath,amssymb}
\usepackage{gensymb}
\usepackage{times}
\usepackage{nicefrac}
\usepackage{amsmath}
\usepackage{xcolor}
\usepackage[normalem]{ulem} 
\usepackage{hyperref}
\usepackage[makeroom]{cancel}

\input{preamble.tex}

\sloppypar

\begin{document}
\title{The Milky Way's circular velocity curve measured using element abundance gradients}

\newcommand{\affcca}{
    Center for Computational Astrophysics, Flatiron Institute, 
    162 Fifth Ave, New York, NY 10010, USA
}

\newcommand{\affcolumbia}{
    Department of Astronomy, Columbia University, 
    550 West 120th Street, New York, NY 10027, USA
}

\newcommand{\affed}{
    Institute for Astronomy, University of Edinburgh, Royal Observatory, Blackford Hill, Edinburgh, EH9 3HJ, UK
}

\newcommand{\affcam}{
    Institute of Astronomy, University of Cambridge, Madingley Road, Cambridge CB3 0HA, UK
}

\newcommand{\affccpp}{
    Center for Cosmology and Particle Physics, Department of Physics, New York University, 
    726 Broadway, New York, NY 10003, USA
}
\newcommand{\affmpia}{
    Max-Planck-Institut f\"ur Astronomie, 
    K\"onigstuhl 17, D-69117 Heidelberg, Germany
}

\newcommand{\affking}{
    Department of Physics, Engineering Physics, and Astronomy, 
    Queen’s University, Kingston, Ontario, Canada
}

\newcommand{\affsurr}{
    School of Mathematics \& Physics, University of Surrey, Guildford, GU2 7XH, UK
}

\author[0000-0003-1856-2151]{Danny Horta}
\affiliation{\affed}

\author[0000-0003-0872-7098]{Adrian~M.~Price-Whelan}
\affiliation{\affcca}

\author[0000-0001-8917-1532]{Sergey E. Koposov}
\affiliation{\affed}
\affiliation{\affcam}

\author[0000-0001-8917-1532]{Jason A. S. Hunt}
\affiliation{\affsurr}

\author[0000-0003-2866-9403]{David~W.~Hogg}
\affiliation{\affcca}
\affiliation{\affccpp}
\affiliation{\affmpia}

\author[0000-0001-5522-5029]{Carrie Filion}
\affiliation{\affcca}

\author[0000-0003-2594-8052]{Kathryn J. Daniel}
\affiliation{Department of Astronomy and Steward Observatory, University of Arizona, 933 N Cherry Avenue, Tucson, AZ 85721, USA }

\correspondingauthor{Danny Horta}
\email{dhorta@roe.ac.uk}

\begin{abstract}\noindent
Spectroscopic surveys now supply precise stellar label measurements such as element abundances for large samples of stars throughout the Milky Way.
These element abundances are known to correlate with orbital actions or other dynamical invariants. 
We present a new data-driven method for empirically measuring the circular velocity curve of the Galaxy that uses element abundance gradients in the plane of radial kinematics. 
We use stellar surface abundances from the \textsl{APOGEE} survey combined with kinematic data from the \textsl{Gaia} mission. Our results confirm the ordered structure of the Milky Way disk in terms of average [Fe/H] and [Mg/Fe] abundance ratios, and suggest that $\langle$[Fe/H]$\rangle$ traces the radial position of stars in the disk, while $\langle$[Mg/Fe]$\rangle$ traces the orbital excursions around this radius.  
Our method uses the radial orbit structure in the Galaxy to enable an empirical measurement of the circular velocity curve, epicyclic and azimuthal frequencies, and kinematic gradients across the Milky Way disk.
From these measurements, we infer a value of the circular velocity curve at the Solar radius of $v_{c,\odot} = 235.3^{+2.8}_{-3.7}$ km s$^{-1}$ using the most constraining abundance ratio, [Mg/Fe]. We also measure the radial and azimuthal frequencies for a circular orbit at the solar radius, $\kappa_{0,R_\odot}=36.9^{+0.8}_{-1.0}$ km s$^{-1}$ kpc$^{-1}$ and $\Omega_{0,R_\odot}=28.5_{-0.1}^{+0.4}$ km s$^{-1}$ kpc$^{-1}$, respectively. These values lead to an estimate of the Oort constants of $A = 16.5^{+0.1}_{-0.1}$ km s$^{-1}$ kpc$^{-1}$ and $B=-11.9^{+0.1}_{-0.3}$ km s$^{-1}$ kpc$^{-1}$. We measure the radial acceleration at the Solar radius to be $(\frac{\partial \Phi}{\partial R})_{\odot} = a_{R_\odot}=7.0^{+0.2}_{-0.1}$ pc Myr$^{-2}$.

\end{abstract}

\section{Introduction}

Understanding the properties of galaxies is pivotal to unravel the making of the Universe. A key tool used to understand the structure and evolution of galaxies is dynamics, the study of the motion of a large number of point masses (typically stars) orbiting under the influence of their mutual self-gravity and the pull of dark matter \citep{Binney2008}. However, inferring the orbital and structural evolution of galaxies is a non-trivial task. As we only observe their stellar populations in a snapshot in time, the study of galactic dynamics must use theoretical frameworks built on statistical physics to gain insights about galaxy formation processes. Furthermore, for galaxies outside our own, we typically are not able to resolve their stellar populations individually, and are thus restricted to making general measurements of their structure. 

Although restrictive of its complete view, our position within the Galaxy ---the Milky Way--- enables resolved measurements of the positions and motions of its stars that is currently unrivaled by any other galaxy in the Universe. Thanks to the advent of \textsl{Gaia} \citep{Gaia2016, Gaia2022}, we now have precise measurements of the positions and motions for over $\gtrsim1.5$ billion sources from astrometric solutions. This vast catalogue has provided the necessary data that has revolutionized our understanding of the Galaxy's structure, providing fresh measurements of its mass distribution \citep[e.g.,][]{Gaia2018_disc,Widmark2021, Widmark2022,Gaia2023_disc, Gaia2023_drimmel,Hunt2025,Ibata2024} and the amount of dark matter it contains \citep[][]{Salomon2020,DeSalas2021, Ou2024,Lim2025, Putney2025}; it has also provided insight into the structure of the Milky Way's closest satellites \citep[e.g.,][]{Luri2021, Vasiliev2021,Jimenez2023, Jimenez2025}. Moreover, these data have also revealed the extent to which the Galaxy is in disequilibrium \citep[e.g.,][]{Antoja2018, Hunt2022, Hunt2024}, providing further challenges to modeling the structure and evolution of the Galaxy.

In addition to the detailed astrometric information supplied by \textsl{Gaia}, large-scale spectroscopic surveys (\textsl{APOGEE}: \citealp{Majewski2017}; \textsl{DESI}: \citealp{Cooper2023}; \textsl{GALAH}: \citealp{Martell2017}; \textsl{LAMOST}: \citealp{Cui2012}; \textsl{MWM}: \citealp{Kollmeier2017}, among others) provide precise spectroscopic information in the form of line-of-sight (LOS) velocities, stellar parameters, and chemical abundances for a spectrum of element species. Of particular importance has been the supply of detailed element abundance information, as the chemical compositions of stars are a direct tracer of the environment in which they were born \citep{Freeman2002}. Thus, chemical abundance ratios provide a way of tagging groupings of stars with the same origin. This has served particularly useful for studies of the Milky Way stellar halo \citep[e.g.,][]{Hawkins2015,Das2020,Monty2020,Horta2021, Lucey2022,Horta2023, Monty2024}. For the Milky Way disk, element abundance measurements have proven useful for describing its structure \citep[e.g.,][]{Rix2013,Hayden2015, Bovy2016, Mackereth2017,Ness2019,Imig2023, Gaia2023_chemcart, Akbaba2024}.

With the wealth of spectroscopic and astrometric data, there has been a push for new dynamics frameworks that exploit chemical abundance data to map correlations with kinematic and orbital quantities \citep[e.g.,][]{Sanders2015, Price2021, Binney2023, Price2025, Widrow2025}. The details between the different methods vary depending on what kinematic/orbital quantity is used to model, and its associated assumptions, leading to more flexible or more interpretable methods. However, all these chemical-dynamical frameworks rely on the fact that there are element abundance correlations with kinematic/dynamical quantities, thus exploiting the fact that element abundance gradients in the Milky Way (thin) disk trace the orbits of its stars. 
A new method, called Orbital Torus Imaging (OTI) \citep{Price2021, Horta2024_oti,Price2025, Frankel2024}, exploits mono-abundance contours in kinematic space to empirically map the orbits of stars.  Building on these works, in this article we introduce a new framework for empirically measuring the circular velocity curve of the Galaxy using gradients of mean element abundances in kinematic space.

This manuscript is organized as follows: Section~\ref{sec_data} describes the data we use and the selection criteria to obtain our working sample; in Section~\ref{sec_gradients}, we set out to understand the chemical-orbital structure of the Milky Way's low-$\alpha$ disk by examining the correlations of mean [Fe/H] and mean [Mg/Fe] as a function of radial and vertical kinematics; Section~\ref{sec_model} describes the method used in this paper; Section~\ref{sec_results} presents the main body of results from this work, where we specifically show how it is possible to empirically obtain the circular velocity curve of the Milky Way by modelling abundance gradients in the \textsl{APOGEE-Gaia} low-$\alpha$ disk; we close by discussing our results in Section~\ref{sec_discussion} and presenting our conclusions in Section~\ref{sec_conclusions}.

\section{Data}
\label{sec_data}

\begin{figure*}
    \centering
    \includegraphics[width=\textwidth]{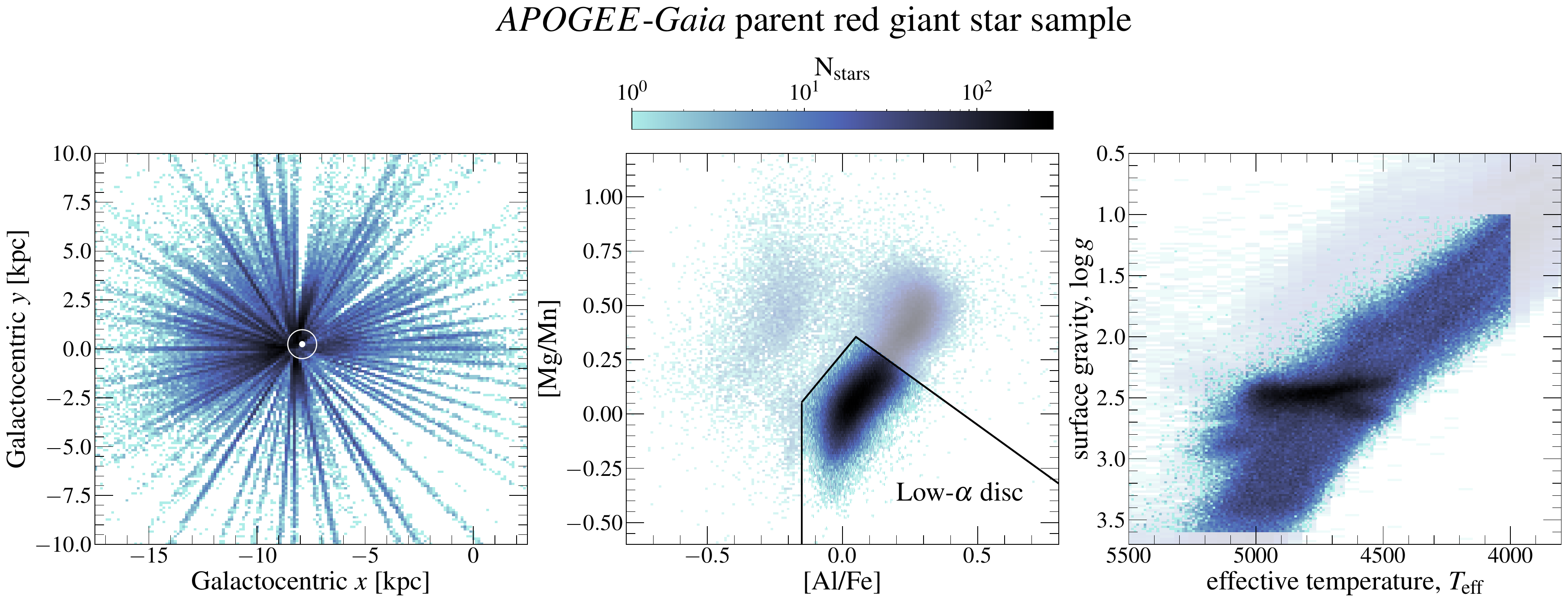}
    \caption{Our parent sample of \textit{APOGEE-Gaia} red giant branch stars in the low-$\alpha$ disk used in this work. The left panel shows a 2D histogram of the positions of the stars in our sample projected onto the Galactic plane (i.e., in Galactocentric [$x$, $y$] coordinates). The approximate position of the Sun is marked with the $\odot$ symbol. The middle panel shows the selection used to define the low-$\alpha$ sequence using a cut in element abundance space, i.e., the [Mg/Mn] vs. [Al/Fe] plane. This chemical cut was performed in order to restrict our sample to stars that are part of the low-$\alpha$ disk, and thus likely part of the Galaxy’s thin disk. The right panel shows the spectroscopic stellar parameters, effective temperature $T_{\mathrm{eff}}$ and surface gravity $\log~g$. In the middle and right panels, the full \textsl{APOGEE–Gaia} sample is shown as the faint background histogram.}
    \label{fig_low_alpha}
\end{figure*}

We use a cross-matched catalog comprised of the final spectroscopic data release (DR17) from the \textsl{APOGEE} survey (\citealp{Majewski2017, SDSSDR17}) and astrometric data from \textsl{Gaia} DR3 \citep{Gaia2022}. \textsl{APOGEE} is a dual-helmisphere survey that collected near-infrared data using two high-resolution spectrographs \citep{Wilson2019} mounted on the $2.5$m telescope at Apache Point Observatory \citep{Gunn2006} in New Mexico, and the Du Pont telescope at Las Campanas Observatory \citep{BowenVaughan1973} in Chile. Stellar parameters and element abundances are determined using the \textsl{ASPCAP} pipeline \citep{Perez2015} based on the \textsl{FERRE} code \citep{Prieto2006}, using the line-lists from \cite{Cunha2017} and \cite{Smith2021}. The spectra were reduced by a customized pipeline \citep{Nidever2015}. All target selection criteria for \textsl{APOGEE} and \textsl{APOGEE-2} are described in \citet{Zasowski2013} and \citet{Zasowski2017}, respectively; details of \textsl{APOGEE}-north can be found in \citet{Beaton2021}, whereas information about \textsl{APOGEE}-south are contained in \citet{Santana2021}.

The \textsl{Gaia} \citep{Gaia2016} mission delivers detailed sky positions, proper motions, parallax measurements, and low-resolution optical spectra for $\approx1.5$ billion stars with apparent magnitudes brighter than (\textsl{Gaia}) $G < 20.7$. Here, we use the astrometric (position, parallax, and proper motion) measurements from \textsl{Gaia} DR3 \citep{Gaia2022}. We also use photo-astrometric distances derived from \textsl{Gaia} data (namely, the \textsl{StarHorse} distance catalogue: \citealp{Anders2022}).

In combination, the LOS velocities from \textsl{APOGEE} and the astrometry from \textsl{Gaia} provide full 6D phase-space information from which the kinematic and orbital properties can be derived. The \textsl{APOGEE} data additionally provide
detailed element abundance measurements for up to $\sim20$
different species spanning different nucleosynthetic channels: the $\alpha$, odd-Z, iron-peak, and $s$-process. 

\subsection{Parent sample}
\label{sec_data_parent}
The parent sample used in this work is shown in Figure~\ref{fig_low_alpha} and is comprised of stars that satisfy the following criteria:

\begin{description}
    \item[Red Giant Branch disk stars] \textsl{APOGEE}-determined atmospheric parameters: effective temperatures and surface gravities of $4,000 < T_{\mathrm{eff}} < 5,500$ K and $1 < \log~g <3.6$, respectively. To ensure we only work with stars targetted in disk fields, we select stars with $(J - K_s) \geq 0.5$.
    \item[Medium signal-to-noise spectra] \textsl{APOGEE} spectral S/N $>50$.
    \item[High-quality derived spectral parameters] \textsl{APOGEE} \texttt{STARFLAG} bit numbers not set to 0, 1, 3, 16, 17, 19, 21, or 22 and \textsl{APOGEE} \texttt{ASPCAPFLAG} bit numbers not set to 23.
    \item[No star clusters] Stars that are not within the \textsl{APOGEE} globular cluster value added catalog \citep{Schiavon2024}.
    \item[Low-$\alpha$ stars] Stars that fall in the low-$\alpha$ sequence of the [Mg/Mn]-[Al/Fe] plane (\citealp{Horta2021}, see Figure~\ref{fig_low_alpha}). These stars are defined as $-0.15<\mathrm{[Al/Fe]}<0.05 \wedge  \mathrm{[Mg/Mn]}>1.5\times \mathrm{[Al/Fe]}+0.28 \vee \\0.05<\mathrm{[Al/Fe]}<0.3 \wedge \mathrm{[Mg/Mn]}>-0.9\times \mathrm{[Al/Fe]}+0.4$.
    \item[High-quality distance estimates] \textsl{StarHorse} distances with $d/\sigma_{d}>5$.
\end{description}
 We use the 6D phase-space information to compute Galactocentric Cartesian coordinates using the \texttt{gala} code \citep{gala}, assuming solar velocity $v_{\odot} = [8.4, 251.8, 8.4]~\mathrm{km ~s}^{-1}$, the distance to the Galactic center $R_0 = 8.275$ kpc \citep{Gravity2021}, and the height from the disk midplane of $z=20.8$ pc \citep{Bennett2019}. 

 Furthermore, for our modeling purposes (Section~\ref{sec_results}), we only use stars within $|z|<0.5$ kpc from the midplane that have low vertical velocities, $|v_z|<30$ km s$^{-1}$, and (cylindrical) radial velocities of $|v_R|<100$ km s$^{-1}$. We choose to cut on kinematics instead of orbital quantities, such as actions, to ensure that we remain potential-independent in our selection criteria. 
 Our selection yields a total of $88,917$ stars.

\begin{figure*}
    \centering
    \includegraphics[width=\textwidth]{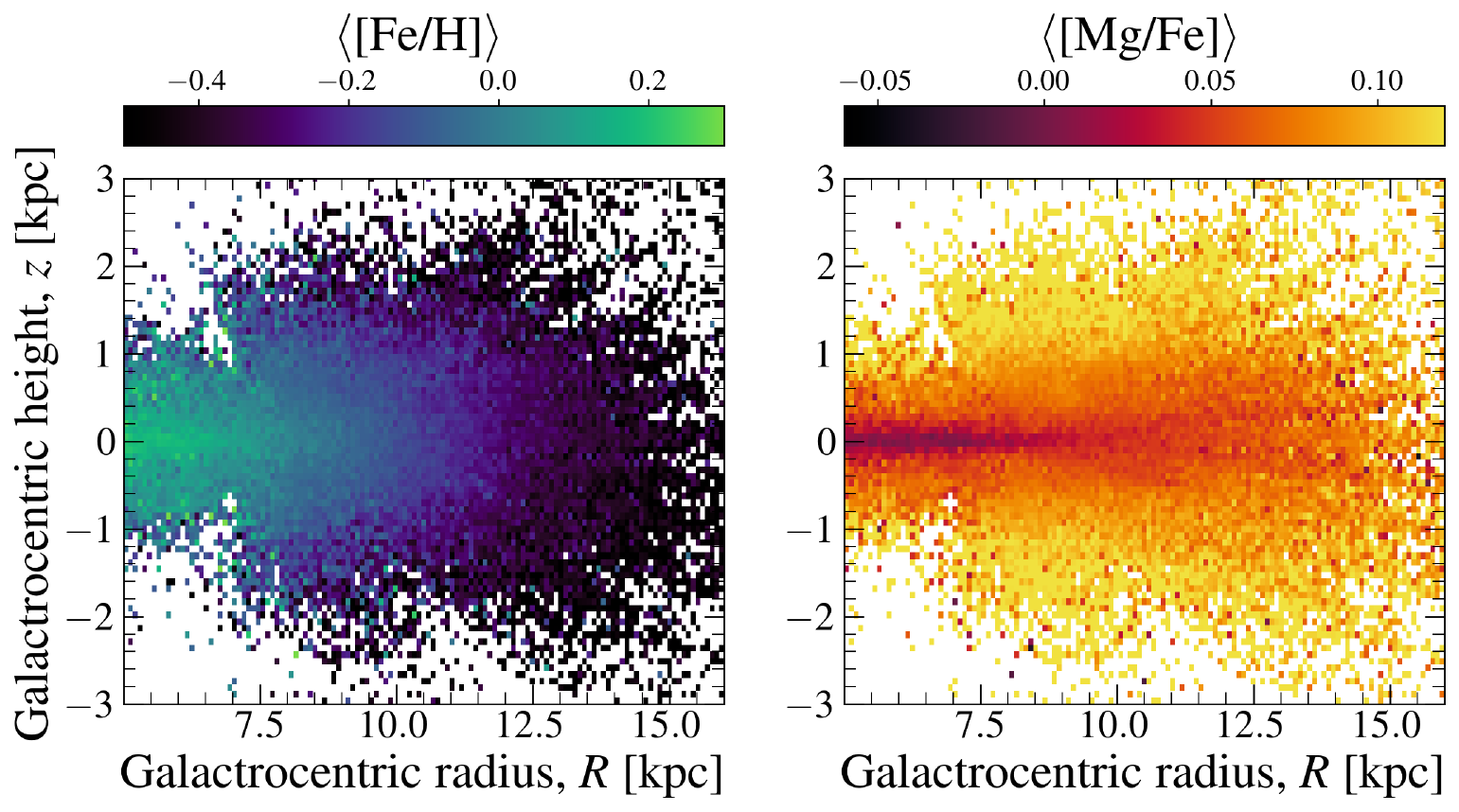}
    \caption{Our low-$\alpha$ sample projected in present day $R$--$z$ (side-on view). Each panel shows the sample binned, where each pixel represents the mean value of [Fe/H] (left) and [Mg/Fe] (right). $\langle$[Fe/H]$\rangle$ shows a gradient that is primarily dependent on Galactrocentric radius, $R$. Conversely, $\langle$[Mg/Fe]$\rangle$ shows a gradient that primarily changes with vertical position, $z$ (although there is some small dependence on $R$). This figure illustrates that the Milky Way low-$\alpha$ disk is well structured both chemically and kinematically for these element abundance ratios.}
    \label{fig_grad_R_z}
\end{figure*}

\section{The structure of the Milky Way's low-$\alpha$ disk}
\label{sec_gradients}

The Milky Way's low-$\alpha$ disk\footnote{The low-$\alpha$ disk is also commonly referred to as the structurally thin disk.} is a well structured component, displaying known correlations between a star's position, velocity, and element abundance ratios. A clear example of this is the radius--[Fe/H]--age relation \citep{Frankel2018}. The Galactic disk is also homogeneous in terms of star formation for most elements examined so far (\citealp{Ness2019, Ness2022}), although there may be deviations in the heavier $s/r$-process elements \citep{Horta2022}. If divided into mono-abundance or mono-age populations (MAPs: \citealp{Bovy2016, Mackereth2017}), there appears to be a continuity between different sub-populations, attesting to its ordered structure. 

The element abundance gradients exhibited in the Milky Way's thin disk vary depending on the dimensions studied. Figure~\ref{fig_grad_R_z} shows the parent sample (as described in Section~\ref{sec_data_parent}) pixelated in cylindrical coordinates radius and height, $R$--$z$, where in the left panel each pixel shows the mean abundance of [Fe/H] and the right panel shows the mean abundance of [Mg/Fe]. The mean [Fe/H] value for stars in the Galactic low-$\alpha$ disk shows a gradient that is primarily dependent on Galactrocentric radius, $R$. Conversely, the mean [Mg/Fe] abundance shows a gradient that primarily changes with vertical position, $z$ (although there is some small dependence on $R$). The [Fe/H] and [Mg/Fe] (i.e., [$\alpha$/Fe]) gradients w.r.t. $R$ and $z$ are a feature of the Milky Way disk that has been known for some time and has been shown extensively in previous works (e.g., \citealp{Cheng2012,Schlesinger2014,Hayden2015, Bovy2016, Mackereth2017,Imig2023}). Nevertheless, it is a key insight and an interesting property of the Galaxy that can be exploited for jointly modeling the elemental abundances and dynamics of the disk. For example, \citet{Price2025} show that it is possible to exploit the gradients seen in vertical kinematics to empirically model the mass distribution in the Galaxy. This is because mono-abundance contours trace orbits with different vertical action (or vertical energy), which foliate the vertical phase space at a given radius in the disk \citep{Price2021}; it is therefore possible to see projections of the orbital tori with mono-abundance contours. 

\begin{figure*}
    \centering
    \includegraphics[width=\textwidth]{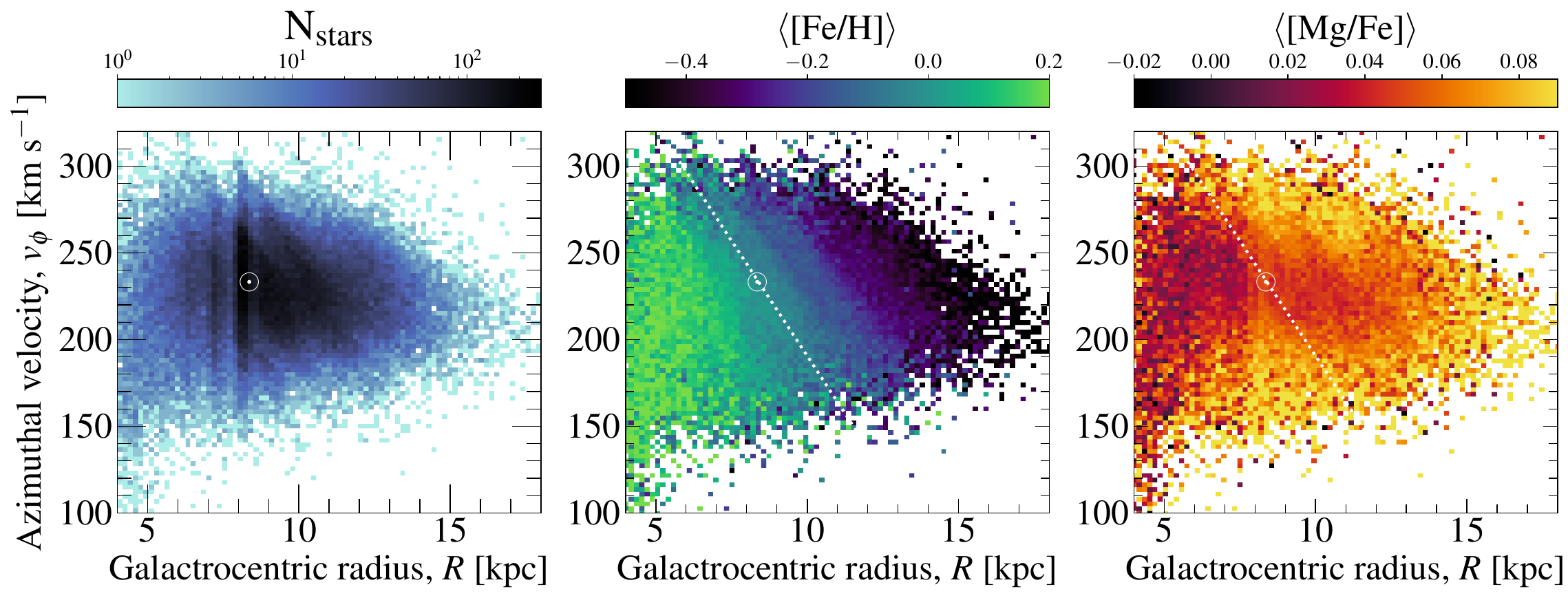}
    \caption{Low-$\alpha$ sample in the $v_\phi$--$R$ plane. The left panel shows a 2D density histogram, the middle panel shows the data pixelated by average [Fe/H], and the right panel shows the data pixelated by average [Mg/Fe]. The left panel illustrates the kinematic structure in the Milky Way's disk; the jagged features (or ridges) are a result of features in the in-plane dynamics of stars, whereas the vertical lines around $R\approx7$-$8$ kpc are caused by the spatial selection function of the \textsl{APOGEE} survey. The middle panel shows how $\langle$[Fe/H]$\rangle$ traces angular momentum, $L_z = R\,v_\phi$, leading to stripes of constant $\langle$[Fe/H]$\rangle$ (e.g., dotted white line). The right panel shows how $\langle$[Mg/Fe]$\rangle$ traces the out-of-plane motions of stars in the low-$\alpha$ disk. Here, the ridges of constant $\langle$[Mg/Fe]$\rangle$ are qualitatively equal to the ones in density. The approximate location of the Sun is marked by the $\odot$ symbol.}
    \label{fig_R_vphi}
\end{figure*}

Figure~\ref{fig_R_vphi} shows the low-$\alpha$ disk sample (i.e., the parent sample with $|z|<0.5$ kpc, $|v_z|<30$ km s$^{-1}$, and $|v_R|<100$ km s$^{-1}$) in the azimuthal velocity versus Galactocentric radius plane, $v_\phi$--$R$, where we have marked the location of the Sun with a $\odot$ symbol. The left panel shows a 2D density histogram, the middle panel shows the data pixelated by average [Fe/H], and the right panel shows the data pixelated by average [Mg/Fe]. The left panel is analogous to Figure~2 from \citet{Antoja2018}, and illustrates the kinematic structure in the Milky Way's low-$\alpha$ disk; the jagged features (or ridges) are a result of features in the in-plane dynamics of stars. The middle panel shows how $\langle$[Fe/H]$\rangle$ traces angular momentum, $L_z = R\,v_\phi$, leading to stripes of constant $\langle$[Fe/H]$\rangle$ (e.g., the dotted white line). This result is the same as the trend we see in the left panel of Figure~\ref{fig_grad_R_z}. The right panel shows how $\langle$[Mg/Fe]$\rangle$ displays a different gradient to $\langle$[Fe/H]$\rangle$, one that instead of tracing $L_z$ traces the out-of-plane (vertical) motions of stars in the low-$\alpha$ disk. Here, the ridges of constant $\langle$[Mg/Fe]$\rangle$ are qualitatively equal to the ones in density. This result further supports the results from Figure~\ref{fig_grad_R_z}, and suggests that the gradients in $\langle$[Fe/H]$\rangle$ and $\langle$[Mg/Fe]$\rangle$ also correlate with stellar orbits (in this case, angular momentum).

\subsection{Element abundance trends in radial kinematics}
\label{subsec_radialcoords}

\begin{figure*}
    \centering
    \includegraphics[width=\textwidth]{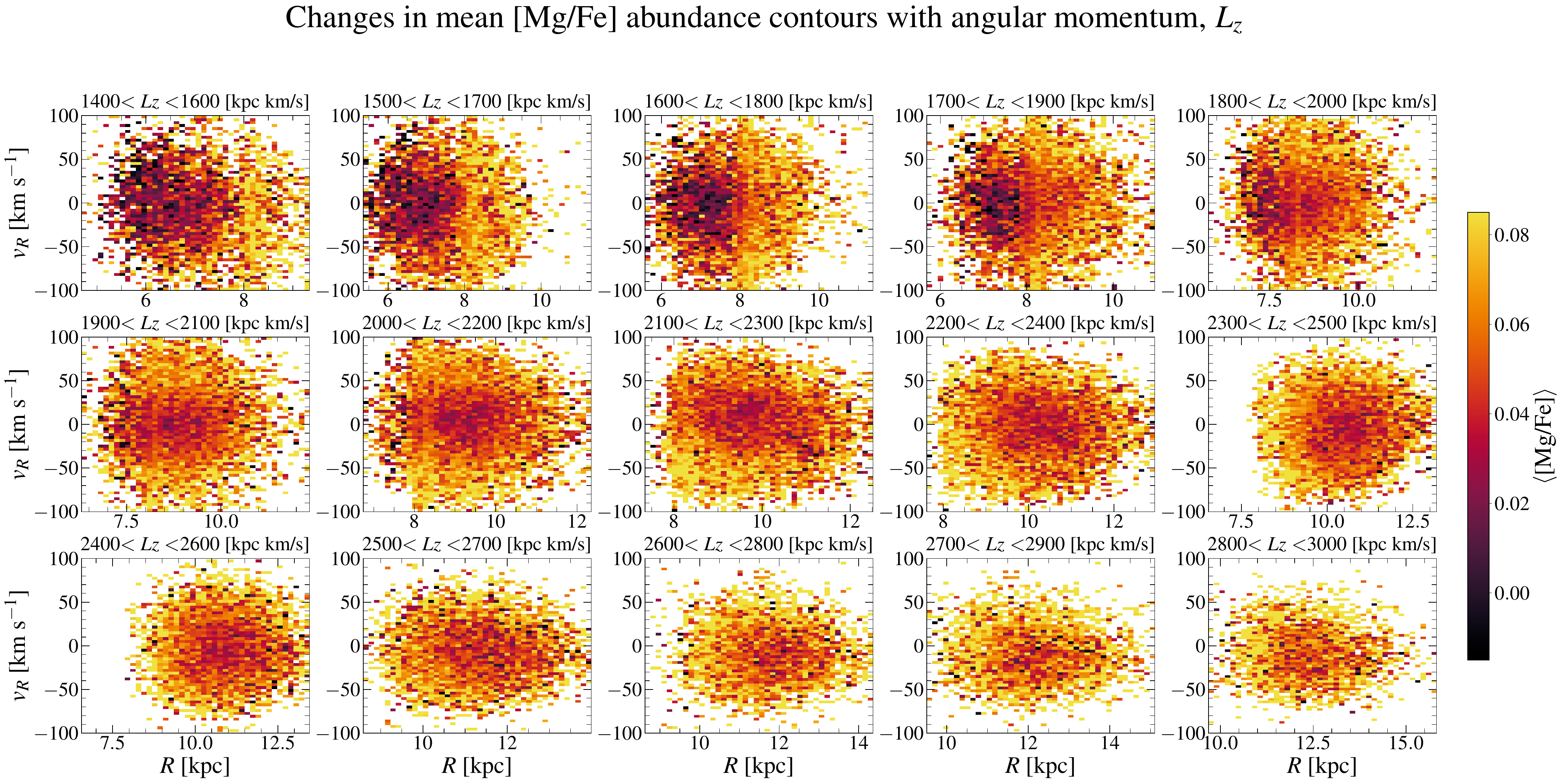}
    \caption{The parent low-$\alpha$ disk sample used in this work illustrated in the plane of radial kinematics ($R$--$v_R$), binned as a function of angular momentum, $L_z = R \, v_\phi$. In each panel, the data is pixelated, where each bin shows the mean value of [Mg/Fe] for all stars in each pixel. In the first panel (top left), there is a stripe of stars with $R\sim8$ kpc across all $v_R$, which appears due to the spatial selection function of the survey and the fact that more stars on higher non-circular orbits at their apocenter will be observed in bins closer to the inner Galaxy. In all bins the distribution of mono-abundance contours (i.e., the gradient with mean [Mg/Fe]) appears to trace a guitar-pic or arrow-head pattern, as expected for the epicyclic motion of stars on one side of the Milky Way disk \citep[see Figure~\ref{fig_contours} or][for example]{Hunt2024}. The shape of this guitar-pic structure however changes across different $L_z$ bins. At lower $L_z$, the distribution of the data is broader in $v_R$ and narrower in $R$; at larger $L_z$, the data show a broader distribution in $R$ and a narrower range in $v_R$. This change in shape is caused by the larger enclosed mass in the inner regions of the Galaxy. At larger radii, where the enclosed mass is lower, stars with the same or even smaller radial velocities reach larger radial excursions. Interestingly, we don't find this guitar-pic pattern for mono-abundance contours when examining the gradients with average [Fe/H] (see Figure~\ref{fig_app_lz_feh}).}
    \label{fig_abun_lz_bins}
\end{figure*}

\begin{figure*}
    \centering

    \includegraphics[width=\textwidth]{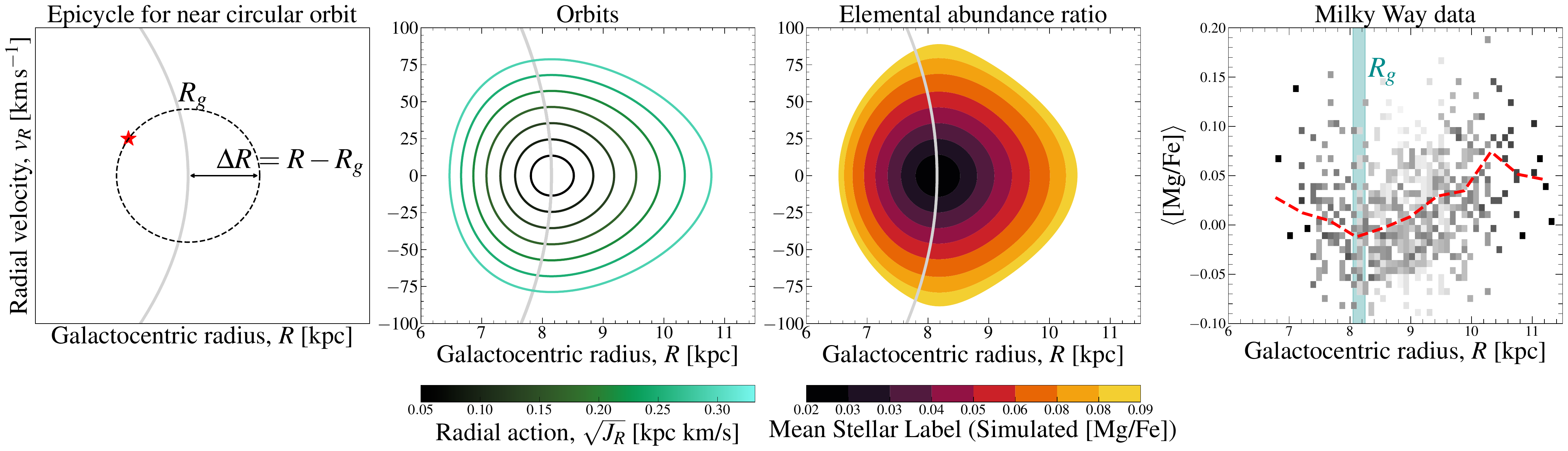}
    \caption{This figure demonstrates that, for stars in a narrow slice of $L_z$, contours of constant element abundance ratio nearly delineate orbits in $\{R, v_R\}$ space, and trace the epicyclic excursions for stars with near-circular orbits in the Milky Way's low-$\alpha$ disk. It also shows how it is possible to empirically measure the circular velocity curve by finding the inflexion point in the element abundance trend. 
    \textit{\textbf{Left}}: Cartoon of the epicyclic motion of a star (red star symbol) in the Milky Way's disk on a near-circular orbit. Here, the solid gray line delineates the guiding-center radius, $R_g$, of a star orbiting around the Galactic center, and the dashed black line delineates the epicyclic motion of that star around its guiding-center radius. The difference between the star's observed position and its guiding-center radius, $\Delta R = R - R_g$, defines the magnitude of the radial excursion of that near-circular orbit.
    \textit{\textbf{Center left}}: This panel shows seven orbits with different values of the radial action $J_R$ computed in a simple Milky Way mass model.
    For all of these orbits, the value of the other actions are set such that the orbits have zero vertical action, $J_z=0$, and a constant azimuthal action set to the solar value, $J_\phi = J_{\phi, \odot}$. Orbits with lower orbital action trace smaller trajectories (darker lines), are more elliptical, and reach smaller maximum excursions from the guiding-center radius, $R_g$. 
We note that this is our only usage of a parameterized Galactic mass model in this work, and this is only for demonstration purposes.
    \textit{\textbf{Center right}}: We paint element abundances onto the simulated stars with a linear dependence on radial action $J_R$ with a slope similar to that of [Mg/Fe] (see the right panel).
    The filled contours then show curves of constant mean element abundance ratio for a simulated population of orbits in the radial phase space, going from smaller average [Mg/Fe] in the center to higher values on the outskirts (similar to what is seen for Milky Way data in Figure~\ref{fig_abun_lz_bins}).
    \textit{\textbf{Right}}: Milky Way \textit{APOGEE-Gaia} data for low-$\alpha$ disk stars in a narrow slice of $|v_R|<10$ km s$^{-1}$ around the Sun ($1900 < L_z< 2050$ kpc km s$^{-1}$), and with $|z|<0.1$ kpc, $|v_z|<10$ km s$^{-1}$. Here, we show a 2D density distribution of the data and its dependence of $\langle$[Mg/Fe]$\rangle$ with $R$, overlaying the running median as a dashed red line. As can be seen from the figure, for stars in a small bin of $L_z$, the relationship between average chemical abundance with $R$ in a thin slice of $v_R$ around $v_R\approx0$ leads to an inflexion point in the mean abundance distribution that aligns with the guiding-center radius for stars in that bin (the vertical shaded line). Since $v_c = L_z/R_g$, this approach provides a way of empirically measuring the Milky Way's circular velocity curve using element abundance gradients.}
    \label{fig_contours}
\end{figure*}

Figure~\ref{fig_abun_lz_bins} shows the parent low-$\alpha$ disk sample split in 15 overlapping bins of angular momentum, $L_z$, each of 200 kpc km~s$^{-1}$ in width. In each panel, we show the dependence of $\langle$[Mg/Fe]$\rangle$ with radial velocity, $v_R$, and Galactocentric radius, $R$ (i.e., the plane of radial kinematics). The gradient of the average value of [Mg/Fe] in each of the $L_z$ bins shows an interesting dependence with $v_R$ and $R$. Stars with low $v_R$ display lower $\langle$[Mg/Fe]$\rangle$ values than stars with higher $v_R$, and stars closer to the median value of $R$ in each $L_z$ bin also present lower $\langle$[Mg/Fe]$\rangle$ than stars at the tails of the distribution. For each bin of $L_z$, this leads to a gradient in $\langle$[Mg/Fe]$\rangle$ that depends on both $v_R$ and $R$. Moreover, tracing the trajectory delineated by pixels with the same $\langle$[Mg/Fe]$\rangle$ (namely, a mono-abundance contour) results in an guitar-pic or arrow-head like structure. This guitar-pic morphology is the same as that expected for near-circular orbits of stars in the Galaxy's disk during their epicycle around their guiding radius, $R_g$ (see Figure~\ref{fig_contours} or \citealt{Hunt2024}). The shape of this guitar-pic structure however changes across different $L_z$ bins. At lower $L_z$, the distribution of the data is broader in $v_R$ and narrower in $R$; at larger $L_z$, the data show a broader distribution in $R$ and a narrower range in $v_R$. This change in shape is caused by the larger enclosed mass in the inner regions of the Galaxy. At larger radii, where the enclosed mass is lower, stars with the same or even smaller radial velocities reach larger radial excursions.

A narrow slice in $L_z$ will select stars in the low-$\alpha$ disk that orbit around a similar $R_g$ (see the middle panel of Figure~\ref{fig_R_vphi}), since $L_z = v_c/R_g$ (here $v_c$ is the circular velocity curve). Thus, for a small bin in $L_z$, the fact that we see a trend in radial kinematics with $\langle$[Mg/Fe]$\rangle$, and that mono-abundance contours delineate the guitar-pic pattern expected for near-circular orbits on their epicyclic motion, implies that the gradient of $\langle$[Mg/Fe]$\rangle$ and the shape delineated by stars with similar average abundance are contouring the (radial) orbits of stars in the low-$\alpha$ disk. In other words, stars with lower radial action, $J_R$, typically have lower $\langle$[Mg/Fe]$\rangle$ when compared to stars with higher $J_R$. For stars with similar $L_z$, this leads to a foliation of radial orbits in the $R$-$v_R$ plane that can be captured with $\langle$[Mg/Fe]$\rangle$, analogous to the foliation seen in the plane of vertical kinematics for stars with different $J_z$ (see Figure 3 in \citealt{Horta2024_oti}). Interestingly, this link between radial orbits and abundances is not seen for $\langle$[Fe/H]$\rangle$ (see Figure~\ref{fig_app_lz_feh} in Appendix~\ref{app_lzbins}), and of all the elements examined, appears strongest for $\langle$[Mg/Fe]$\rangle$ (see Figure~\ref{fig_abun_all_relations} in Appendix~\ref{app_elements}), matching the results from \citet{Horta2024_oti} when examining the plane of vertical kinematics.

\section{Measuring the circular velocity curve from element abundance gradients}
\label{sec_model}

In this Section we introduce the method for empirically measuring the circular velocity curve of the Milky Way's low-$\alpha$ disk by modelling the elemental abundance distribution in the radial phase space.
Conceptually, our measurement method uses the fact that $L_z = R \, v_\phi$, where the Galactocentric radius $R$ and azimuthal velocity $v_\phi$ are close to being observable quantities for stars. 
If we can then empirically measure the guiding-center radius $R_g$ for stars with similar $L_z$, we directly measure the circular velocity $v_c(L_z)$ because $L_z = R_g \, v_c$. 
We therefore need a method to measure the guiding-center radius, which we outline below.

To motivate this, Figure~\ref{fig_contours} illustrates the plane of radial kinematics, $R$ vs. $v_R$, in three panels, and includes a fourth panel to show the relationship of average [Mg/Fe] w.r.t. $R$ for a bin of $L_z$ and a slice in $v_R \approx 0$. In more detail, the left panel shows a cartoon of the epicyclic motion of a star (red star symbol) in the Milky Way's low-$\alpha$ disk on a near-circular orbit. Here, the solid gray line delineates the guiding-center radius, $R_g$, of a star orbiting around the Galactic center, and the dashed black line delineates the epicyclic motion of that star around its guiding-center radius. The difference between the star's observed position and its guiding-center radius, $\Delta R = R - R_g$, defines the magnitude of the radial excursion of that near-circular orbit. The center left panel shows seven orbits with different values of the radial action, $J_R$, computed in a simple Milky Way mass model.  For all of these orbits, the value of the other actions are set such that the orbits have zero vertical action, $J_z = 0$, and a constant azimuthal action set to the solar value, $J_\phi = J_{\phi,\odot}$. Orbits with lower orbital radial action trace smaller trajectories (darker lines), are more circular, and reach smaller maximum excursions from the guiding-center radius (i.e., small $\Delta R)$. We note that this is our only usage of a parameterized Galactic mass model in this work, and this is only for demonstration purposes. The center right panel shows the dependence of average [Mg/Fe] in bins of radial kinematics. Here, we paint element abundances onto the simulated orbits using a linear dependence on vertical action $J_R$ with a slope similar to that of [Mg/Fe] (see the right panel). The filled contours then show curves of constant mean element abundance ratio for a simulated population of orbits in the radial phase space, going from smaller average [Mg/Fe] in the center to higher values on the outskirts. This panel is created for illustration purposes only to highlight the mono-abundance contours seen in Figure~\ref{fig_abun_lz_bins}. Finally, in the right panel we show Milky Way \textit{APOGEE-Gaia} data for low-$\alpha$ disk stars in a narrow slice of $|v_R|<10$ km s$^{-1}$ around the Sun ($1900 < L_z< 2050$ kpc km s$^{-1}$), and with $|z|<0.1$ kpc, $|v_z|<10$ km s$^{-1}$. Here, we show a 2D density distribution of the data and its dependence of $\langle$[Mg/Fe]$\rangle$ with $R$, overlaying the running median as a dashed red line. As can be seen from the figure, for stars in a small bin of $L_z$, the relationship between average chemical abundance with $R$ in a thin slice of $v_R$ around $v_R\approx0$ leads to an inflexion point in the mean abundance distribution that aligns with the guiding-center radius for stars in that bin (the vertical shaded line).

Figure~\ref{fig_contours} is an illustration of the dynamics contained in Figure~\ref{fig_abun_lz_bins}, and is only for demonstration purposes. For stars in a narrow bin of angular momentum, the contours of mean [Mg/Fe] abundance trace orbits with different radial action magnitudes in the plane of radial kinematics. Importantly, the saddle or inflection point in the trend of average abundance (in the case of $\langle$[Mg/Fe]$\rangle$, it is the minimum value) defines the point in the $R$--$v_R$ plane that is the center of the orbital distribution, and therefore aligns with $R = R_g$ and $v_R = v_{R,0}$. Thus, in a given bin of angular momentum, by identifying the radius at which the inflection point in the abundance gradient occurs it is possible to empirically measure the guiding center radius, and by default, the value of the circular velocity curve at that radius (since $v_c = L_z / R_g$). 

To measure the inflection point in the mean abundance trend, one could take a slice around $v_R \approx 0$ (namely, at $\approx \Delta R_{\mathrm{max}}$) and model the trend of $\langle$[Mg/Fe]$\rangle$ with $R$. The saddle point in the distribution will then align with the point where $R = R_g$ (see the right panel of Figure~\ref{fig_contours}). Another approach one could take is to create a generative model that can map the distribution of mean abundance in the plane of radial kinematics that includes a centroid position as a free parameter in the model, which then one would optimise to determine the radius at which the inflection point occurs. In this work, we repurpose the \texttt{TorusImaging}\footnote{\url{https://github.com/adrn/TorusImaging}.} model software \citep{Price2025}, and map the abundance gradient of mean [Mg/Fe] in the plane of $R$ vs $v_R$ for stars in small bins of $L_z$. This model includes as a free parameter the position of the centroid in $R$ and $v_R$ space. Using this model, we will measure empirically the Milky Way's circular velocity curve in small bins of $L_z$ using element abundances. 

Before proceeding onto performing these measurements, we first describe our underlying assumptions and lay down the analytical formulation. We also assess how well our method works by testing it on realistic simulations of a Milky Way disk (see Appendix~\ref{app_simulations}).

\subsection{Underlying assumptions}
\label{sec_assumptions}
The core idea of the method used in this work is that any stellar label (e.g., element abundance ratios) that correlates intrinsically with the orbital properties of stars in the Galaxy can be used in place of dynamical quantities to ``image'' the orbits. For this to be possible, the element abundance distribution should be constant along an orbit, and stars at different phases along an orbit should have the same (average) element abundance distribution; this enables one to use stellar labels to trace the orbits by finding curves of constant label value.

Our ability to measure the circular velocity curve of the Milky Way using element abundances relies on a set of assumptions described below:
\begin{description}
    \item[Axisymmetric] The gravitational potential that the stars orbit in is axisymmetric and smooth. This allows us to integrate across all azimuthal angles to model stars in the plane of radial kinematics in bins of angular momentum.
    \item[Near circular orbits] We require that stars have low eccentricity (or close to zero radial action $J_R\rightarrow 0$) and have similar values of the $z$-component of the angular momentum $L_z$ (or azimuthal action $J_\phi$). In order to get close to this, we have restricted our disk sample to moderate-to-low radial velocities (or radial action), $|v_R|<100~\kms$ and vertical excursions, $|z|<0.5$ kpc and $|v_z|<30$ km s$^{-1}$, and ensure that we model stars in narrow bins of angular momentum, $L_z$ (and thus have small radial excursions).
    \item[Phase-mixed] The stellar distribution function in radial kinematics, $f(R, v_R)$, in any narrow bin of angular momentum is phase mixed. This assumption is the key assumption; it is what ensures that contours of constant element abundance will correspond to curves of constant orbital actions (or other invariants).
    \item[Abundance-independent selection] Our model depends on abundance distributions conditioned on phase-space position. The measurements made by our method are not biased by the spatial selection function provided that every phase-space location was observed using a homogeneous selection in magnitude and color. For the case of the disk field pointings from \textsl{APOGEE} used in this work, that are targetted using $(J - K_s)_0 \geq 0.5$ mag and $H \geq 10$ mag, this condition is satisfied. Moreover, we further ensure that the stars we model satisfy this condition by solely using stars on the red giant branch, with $1 <\log~g<3.6$.
    \item[Good abundance measurements] We assume that the abundance measurements of different stars at different distances, different luminosities, and different evolutionary phases are all consistent with one another. In particular, we assume that the measured surface abundances we use during modeling do not vary significantly (for either observational or physical reasons) with evolutionary phase.
    \item[Negligible measurement uncertainties] When fitting our model below, we always bin our stellar data into small bins of radial phase-space coordinates ($\delta~R,\delta~v_R$). We assume that most measurements of Galactocentric radius and velocity for stars in our sample are more precise than our adopted bin sizes, so we ignore measurement uncertainties on the kinematic quantities. Similarly, we also assume that measurements of the stellar labels are precise enough to be able to estimate the mean of a sample of stars in $R$ vs $v_R$ space.
\end{description}

\subsection{Modeling the mean abundance gradients in radial kinematics}

We adapt the Orbital Torus Imaging (OTI) framework defined in \citet{Price2025} and used in \citet{Horta2024_oti} to model mean abundance contours in the radial phase space $(R, v_R)$ rather than the vertical plane $(z, v_z)$. 
The OTI framework enables modeling the element abundance distribution, or moments of the abundance distribution, as a function of location in phase space. 
As with our past work, we use the mean [Mg/Fe] abundance ratio in $(R, v_R)$, and use OTI to model mono-abundance contours as Fourier-distorted ellipses characterized by a projected distance $\tilde{r}_R$ and angle $\tilde{\theta}_R$ in this phase space, centered on a reference position and velocity (i.e., the centroid).
Here, we also model the radial phase space in bins of angular momentum component $L_z$.
In the vertical case, the reference position corresponds to the location of the midplane.
Here, in the radial phase space for a bin of $L_z$, the reference position corresponds to the guiding-center radius $R_g$.
The elliptical phase-space radius and angle for the radial phase-space are therefore defined as
\begin{align}
\label{eq_tildaR}
    \rRp &= \sqrt{(R - R_g)^{2}\,\kappa_{0} + (v_R - v_{R, 0})^{2}\,\kappa_{0}^{-1}},\\
    \thRp &= \tan^{-1}\left(\frac{v_R - v_{R, 0}}{R - R_g} \, \kappa_{0}^{-1}\right)
\end{align}
where $v_{R, 0}$ is the centroid in velocity, which should be zero for an equilibrium system but accounts for any bulk motion in the radial kinematic data. 
For this work, the only model parameter we use later for interpretation is the inferred guiding-center position $R_g$.

In order to infer the guiding-center radius for a given bin in $L_z$, we fit a full OTI model of the radial phase space, which works by modeling the shapes of level sets of $\langle \mathrm{[X/Y]} \rangle$ in $(R,v_R)$ space. We do this by parameterizing the shapes of the contours as a low-order Fourier distortion away from ellipses of constant axis ratio (i.e., constant frequency).
We therefore assume that the mean element abundance values, $Y = \langle \mathrm{[Mg/Fe]} \rangle$, are only a function of a Fourier-distorted elliptical radius \rR,
\begin{equation}
    Y = Y(\rR)
\end{equation}
which is related to the elliptical radius defined in Equation~\ref{eq_tildaR} by
\begin{equation}
\label{eq_rR}
    \rR = \rRp \, \left[1 + \sum_m^{\{1, 2, 3, 4\}} e_m(\rRp) \, \cos(m\thRp) \right] \quad ,
\end{equation}
where here we consider both even and odd Fourier terms, unlike the vertical case.
In the radial phase space, orbits are not symmetric to reflections across the guiding radius because of the angular momentum barrier term in the radial dynamics.
As with \citet{Horta2024_oti}, we use spline functions to represent the distortion functions $e_m(\rRp)$ and the function $Y(\rR)$, but for this work we only use and interpret the inferred guiding-center radius $R_g$ and all other parameters are treated as nuisance parameters.

\subsection{Measuring the circular velocity curve using element abundance trends}
\label{sec_rotcurve}
Our method enables us to empirically measure the circular velocity curve without the need to assume a form of the Galactic potential or integrate orbits. This is because we can directly measure the guiding-center radius with our model when modeling stars in a narrow bin of angular momentum $L_z$. Given the value of the angular momentum bin center $L_z$ and the inferred guiding radius $R_g$, the circular velocity is simply
\begin{equation}
\label{eq_vc}
    v_c = L_z/R_g \quad .
\end{equation}
Therefore, by selecting stars in the low-$\alpha$ disk in small bins of $L_z$, and measuring $R_g$ with OTI, we can directly measure the circular velocity curve of the Galaxy using a data-driven, stellar density-independent, and potential-independent method.

With knowledge of the circular velocity curve, it is possible to derive the epicyclic or radial frequency, $\kappa$, in a bin of angular momentum using the epicycle approximation:
\begin{equation}
\label{eq_kappa_omega}
    \kappa^2(L_z) = \Big(R\frac{d\Omega^2}{dR} +4\Omega^2\Big)_{L_z} \quad ,
\end{equation}
since the azimuthal frequency, $\Omega$, is equal to $\Omega = v_c/R$.

\section{Results}
\label{sec_results}
In this section we present the fits to the data, our measurements of the Milky Way circular velocity curve, and present derived quantities such as the radial and azimuthal frequencies and kinematic gradients (namely, the Oort constants). Table~\ref{tab:results} includes the estimates we obtain for stars at the Solar radius.

\begin{figure*}
    \centering
    \includegraphics[width=0.7\textwidth]{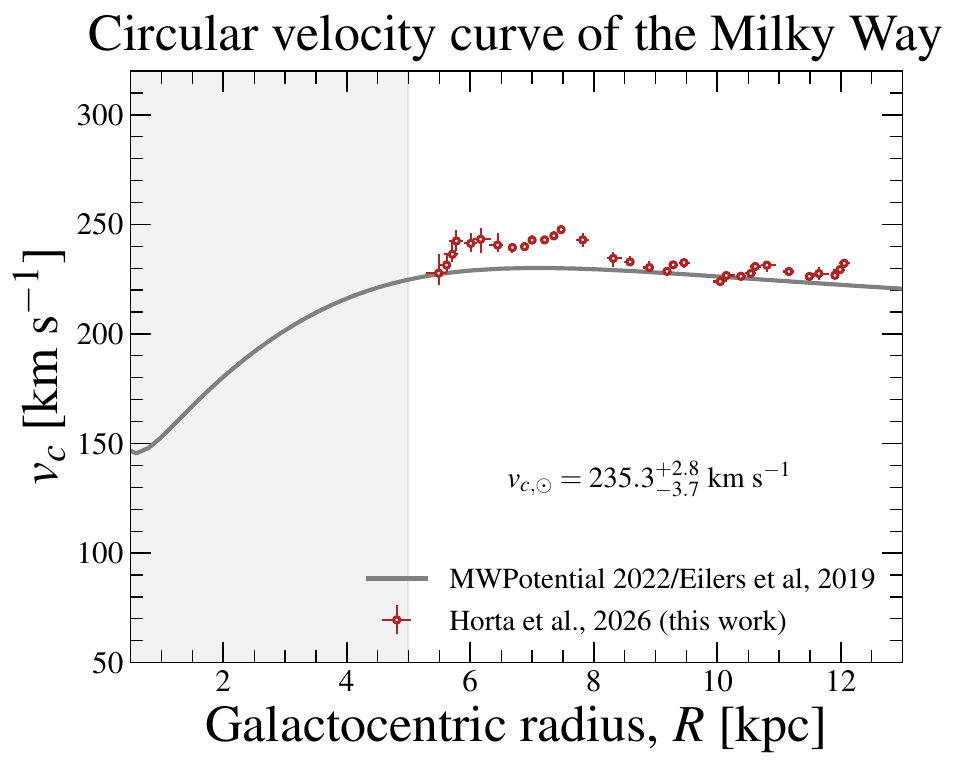}
    \caption{Measured Galactic circular velocity curve as a function of Galactocentric radius, $R$, computed using our method (red) and contrasted against the one predicted by the \texttt{MWPotential2022} model (solid gray line) using the \texttt{gala} package \citep{Price2017}, that itself is made to fit the measurements from \citet{Eilers2019}. The scatter points for our measurements show the median value from 100 bootstrap with replacement realizations of the of $v_R, R-R_g$, [Mg/Fe], and $\sigma_{\mathrm{[Mg/Fe]}}$ distribution in each angular momentum bin modeled, whereas the error bars show the [16$^{\mathrm{th}}$, 84$^{\mathrm{th}}$] percentiles. Here, the radius values, $R$, are equal to the $R_g$ for that bin in $L_z$, since these measurements are computed at the saddle point of the abundance distribution, where $R = R_g$ (see Figure~\ref{fig_contours}). The shaded region at $R<5$ kpc demarks the region of the disk we choose to exclude due to influences of the Galactic bar. Our measurements of the circular velocity curve are in agreement with those from \citep{Eilers2019} between $8<R<12$ kpc, but are in slight disagreement in the inner disk. However, this disagreement is small ($<10\%$), with a maximum difference of $\approx20$ km s$^{-1}$. }
    \label{fig_vc}
\end{figure*}

\subsection{Model setup}
Our Orbital Torus Imaging model is set up using the following arrangement: we compute $L_z$ for all stars using $L_z = R\,v_\phi$ and bin the data into 32 overlapping 100 kpc km s$^{-1}$ ($\approx0.5$ kpc) wide bins, each separated by 50 kpc km s$^{-1}$ ($\approx0.25$ kpc), spanning $L_{z} = [1200,2750]$ kpc (roughly from $5 < R < 12$ kpc). We choose this $L_z$ range to ensure we are modeling a radial range for which we have good data coverage so that our method is able to fit the data well. For each bin, we model the mean [Mg/Fe] abundance gradient in the plane of $v_R$ vs $R$. We choose to model average [Mg/Fe] over other abundance ratios as $\langle$[Mg/Fe]$\rangle$ shows the tightest relationship with radial kinematics (see Figure~\ref{fig_abun_all_relations} in Appendix~\ref{app_elements}). To perform the modeling, we bin the data in each angular momentum bin into 61 pixels per side, spanning from $-3<R-R_{\mathrm{mean}} < 3$ kpc and $-100 < v_R<100$ km s$^{-1}$. We impose eight knots in the label spline function $Y(\rR)$, and $\{8,12,4,4\}$ knots in the spline for the $m=\{1,2,3,4\}$ moments in the Fourier expansion, respectively; we also impose values at the knots for the $m=\{1,3\}$ moments to be negative so that the guitar-pic feature in the $R$ vs $v_R$ plane can be captured. We initialize the Fourier expansion coefficient knots in logarithmic space to $0.05$, except for the $m=2$ moment knots, which we initialize to $-0.5$ to encourage the model to prioritize ``flattening'' in the $R$ vs $v_R$ space.

\subsection{The Galaxy's circular velocity curve measured with element abundances}

Figure~\ref{fig_vc} shows our resulting measurements of the Milky Way circular velocity curve in red compared to the \texttt{MWPotential2022} model from \texttt{gala} \citep{Price2017}, that itself is made to match the measurements of the circular velocity curve from \citet{Eilers2019}. We compute the uncertainties by bootstrapping with replacement the distribution of $v_R, R$, [Mg/Fe], and $\sigma_{\mathrm{[Mg/Fe]}}$ in each angular momentum bin modeled 100 times, and then take the median as our measurement, and [$16^{th},84^{th}$] percentiles as our upper and lower uncertainties. We note that the radius values, $R$, shown in this Figure are equal to the $R_g$ for that bin in $L_z$, since these measurements are computed at the saddle point of the abundance distribution, where $R = R_g$ (see Figure~\ref{fig_contours}).

Our measurements of the circular velocity curve appear to agree remarkably well for most radii with the estimates from \citep{Eilers2019}, but appear to be in slight disagreement for small radii ($5<R<7.5$), reaching a difference of up to $\approx20$ km s$^{-1}$ ($<10\%$). At the Solar radius, we measure $v_{c,\odot} = 235.3^{+2.8}_{-3.7}$ km s$^{-1}$.

\begin{figure}
    \centering
    \includegraphics[width=\columnwidth]{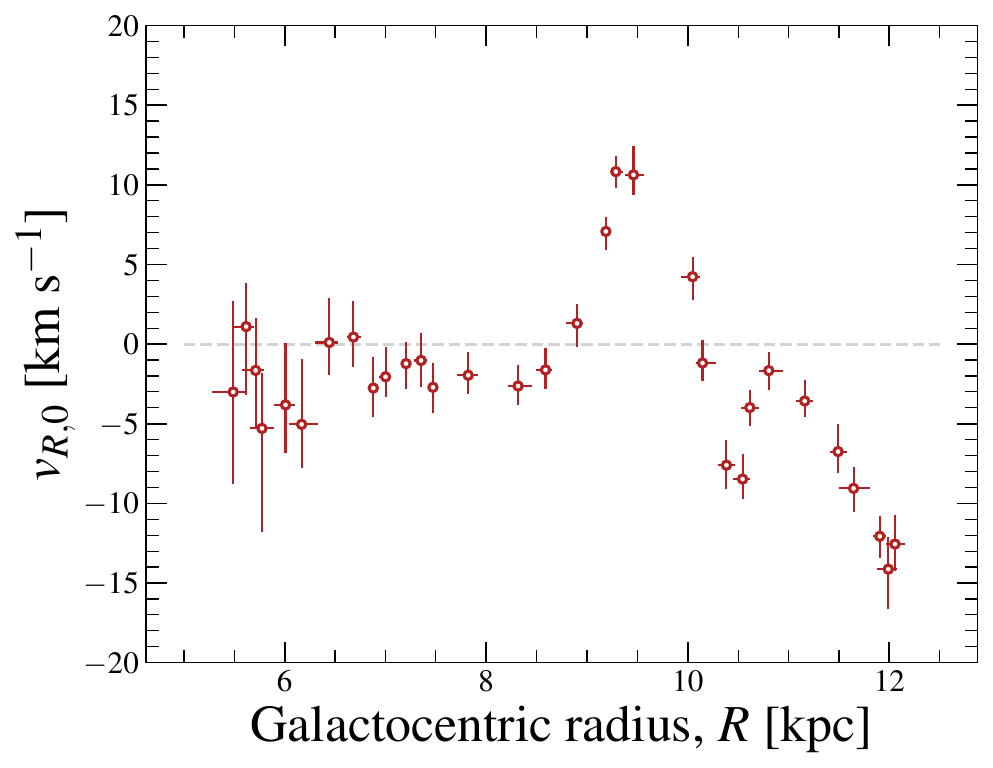}
    \caption{Measurements of the mean radial velocity estimated using our method as a function of Galactocentric radius. The values of $v_{R,0}$ are small but non-zero across the majority of $L_z$ bins modeled. At $R\approx9.5$ kpc, the mean radial velocity increases to a value of $v_{R,0}\sim11$ km s$^{-1}$, after which it drops to a value of $v_{R,0}\sim-15$ km s$^{-1}$ at $R\approx12$ kpc.}
    \label{fig_vr0}
\end{figure}

Interestingly, our results appear to show some ``jumps'' in the $v_c$ profile, and some ``gaps'' in radius; the latter especially around $R\approx9.5$ kpc. These features likely arise from our method capturing dynamical disequilibrium effects. For example, if the radial velocities of stars in that angular momentum bin are non-zero, this indicates that there is some level on dynamical disequilibrium. This could then impact our measurement of $v_c$, as well as the translation from centroid of the element abundance distribution in $R$ to guiding-center radius, thus biasing the average value to lower/higher $R$. Since our method returns an estimate of the mean radial velocity in each $L_z$ bin, we have checked to see if this can be the cause for the gaps seen in $R$. Figure~\ref{fig_vr0} shows the measured $v_{R,0}$ at each radius, and illustrates how these values are non-zero across the majority of bins modeled. Specifically, at the exact location of the most prominent gap in $R$ is at $R\approx9.5$ kpc, we find that the mean radial velocity is high ($v_{R,0}\approx11$ km s$^{-1}$, see also the results from \citealt{Friske2019} and \citealt{Hunt2020}). We discuss the effect of possible dynamical disequilibrium features in more detail in Section~\ref{sec_discussion_diseq} and in Appendix~\ref{app_simulations_live}, where we test our method using a live hydrodynamical simulation.

Lastly, we note that our measurements of $v_c$ are unaffected by the assumed distance from the Sun to the Galactic Center, since the uncertainty reported in \citet{Gravity2021} (namely, $\pm9.3$ pc) is much smaller than the typical uncertainty on $R$ in each $L_z$ bin modelled ($>40$ pc).

\subsection{Epicyclic and azimuthal frequencies}

\begin{figure*}
    \centering
    \includegraphics[width=\textwidth]{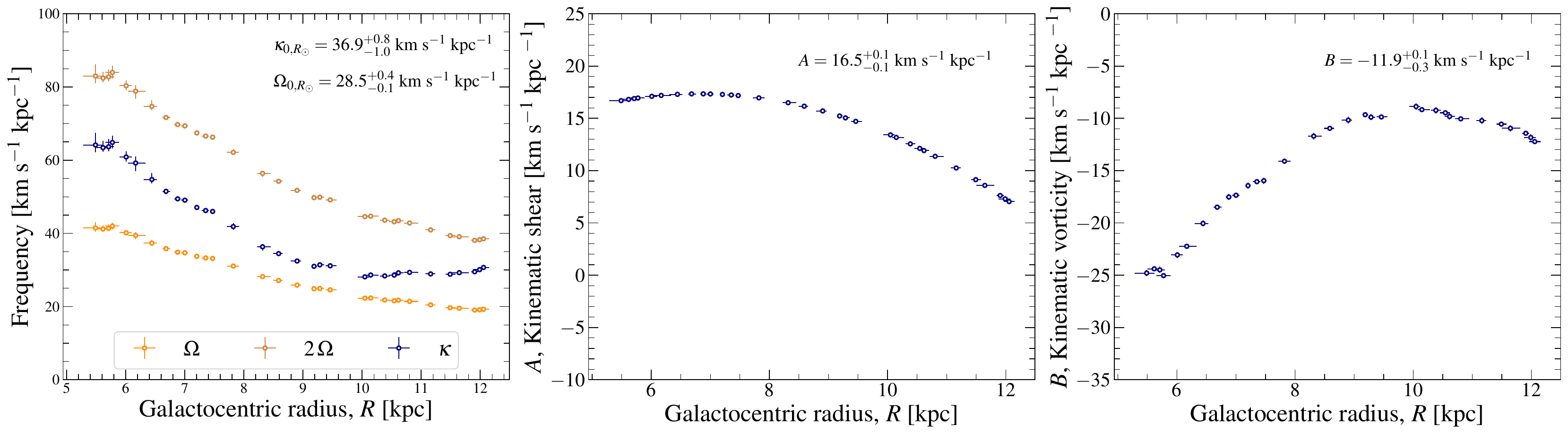}
    \caption{ \textit{\textbf{Left}}: Our empirical measurements of the epicyclic frequency ($\kappa$, navy) and azimuthal frequency ($\Omega$, orange) determined using our measurements of the circular velocity curve. For stars on near circular orbits in the Milky Way's disk, one would expect both the epicyclic and azimuthal frequencies to decrease with increasing $R$, whereby $\kappa$ would adopt a value between $\Omega \lesssim \kappa \lesssim 2\Omega$, as seen from our results for $R\lesssim12$. However, while the azimuthal frequency and $\kappa$ decrease monotonically, we do observe some that at large Galactocentric raii, $\kappa$ plateaus around $\kappa\approx30$ km s$^{-1}$ kpc$^{-1}$. At the Solar radius, we measure a value of $\kappa_{0,R_\odot} = 36.9^{+0.8}_{-1.0}$ km s$^{-1}$ kpc$^{-1}$ and $\Omega_{0,R_\odot} = 28.5_{-0.1}^{+0.4}$ km s$^{-1}$ kpc$^{-1}$. \textbf{\textit{Center}}: Measurements of kinematic shear. In the inner disk ($R\approx6$), we measure a value of $\approx18$ km s$^{-1}$ kpc$^{-1}$, that then drops to $\approx7$ km s$^{-1}$ kpc$^{-1}$ at $R\approx12$. At the Solar radius, we measure a value of the Oort constant of $A=16.5^{+0.1}_{-0.1}$ km s$^{-1}$ kpc$^{-1}$. \textbf{\textit{Right}}: Measurements of kinematic vorticity. In the inner disk ($R\approx6$), we measure a value of $\approx-25$ km s$^{-1}$ kpc$^{-1}$. This value increases to $\approx-10$ km s$^{-1}$ kpc$^{-1}$ at $R\approx10$ and then turns around to drop to $\approx-12$ km s$^{-1}$ kpc$^{-1}$ in the outer regions ($R\approx12$). At the Solar radius, we measure a value of the Oort constant of $B=-11.9^{+0.1}_{-0.3}$ km s$^{-1}$ kpc$^{-1}$.  
    }
    \label{fig_freq}
\end{figure*}

The left panel of Figure~\ref{fig_freq} shows the resulting values of the epicyclic frequency ($\kappa$, blue) at each corresponding Galactocentric radius for the angular momentum bins we fit, obtained using Equation~\ref{eq_kappa_omega} and the measurements of the circular velocity curve from Figure~\ref{fig_vc}. For the latter, we determined the azimuthal frequency, $\Omega = v_c/R$, at each radius and fit the values of $\Omega$ as a function of $R$ with a second order polynomial; we then computed $d\Omega^2/dR = 2\Omega~ d\Omega/dR$ by numerical differentiation using finite differences. In this figure, we also illustrate the value of the azimuthal frequency in orange, $\Omega= v_c/R$, at these radii, computed using the $v_c$ values we obtain from our fitting procedure. For stars on near circular orbits in the Milky Way's disk, one would expect both the epicyclic and azimuthal frequencies to decrease with increasing $R$, whereby $\kappa$ would adopt a value between $\Omega \lesssim \kappa \lesssim 2\Omega$\,\footnote{In an axisymmetric distribution, the decrease of the azimuthal and epicyclic frequencies with radius would never decline faster than the Kepler falloff, $\Omega \propto R^{-3/2}$, that would yield $\kappa=\Omega$.}. This is what we find from our results in Figure~\ref{fig_freq}, where $\kappa$ drops from a value of $\approx60$ km s$^{-1}$ kpc$^{-1}$ at $R\approx6$ kpc to a value of $\kappa \approx30$ km s$^{-1}$ kpc$^{-1}$ at $R \approx 12$ kpc. Across all radii our estimates of $\kappa$ remain within $\Omega$ and $2\Omega$. At the Solar circle, we estimate a value of $\kappa_{0,R_\odot}=36.9^{+0.8}_{-1.0}$ km s$^{-1}$ kpc $^{-1}$ and $\Omega_{0,R_\odot} = 28.5_{-0.1}^{+0.4}$ km s$^{-1}$ kpc $^{-1}$. Interestingly, while the epicyclic frequency decreases monotonically with $R$, we find that at large radii ($R\gtrsim11$) the value of $\kappa$ plateaus at $\kappa\approx30$ km s$^{-1}$ kpc$^{-1}$.

\subsection{Measurements of shear and vorticity in the Milky Way disk and the Oort constants}
 With knowledge of $\kappa$ and $\Omega$, it is possible to determine the Oort constants $A$ and $B$ using stars at the Solar radius. Rearranging Eq\,$3.84$ from \citet{Binney2008} leads to
\begin{align}
    A = -\frac{\kappa^2}{4\Omega} + \Omega \\
    B = -\frac{\kappa^2}{4\Omega}.
\end{align}

Although initially proposed by \citet{Oort1927a} to be fixed parameters, the values of $A$ and $B$ are only constant at the Solar radius. Across a large range of Galactocentric radii, the values of $A$ and $B$ describe the kinematic gradients with respect to the in-plane motions of stars in the Milky Way disk. Thus, $A$ and $B$ are only constants for the region of the Galaxy surrounding the radius at which the corresponding $\kappa$ and $\Omega$ values are computed at. More specifically, $A$ provides a measure of the shear in the disk, whereas $B$ provides a measure of the vorticity.

The center and right panels of Figure~\ref{fig_freq} shows our measurements of kinematic shear and vorticity at each radii, respectively, computed using our measured $\kappa$ and $\Omega$ from the left panel. For kinematic shear, our estimates vary from $A\approx17$ km s$^{-1}$ kpc$^{-1}$ at $R \approx 6$ kpc to $A\approx7$ km s$^{-1}$ kpc$^{-1}$ at $R \approx 12$ kpc. For kinematic vorticity, our estimates yield $B\approx-25$ km s$^{-1}$ kpc$^{-1}$ at $R \approx6$ kpc and $B\approx-12$ km s$^{-1}$ kpc$^{-1}$ at $R\approx12$ kpc. 

For stars at the Solar radius ($R \approx8.275$ kpc), we infer a value of the Oort constants of $A=16.5^{+0.1}_{-0.1}$ km s$^{-1}$ kpc$^{-1}$ and $B = -11.9^{+0.1}_{-0.3}$ km s$^{-1}$ kpc$^{-1}$. These values are approximately equal to previous measurements \citep[$A\approx15$ km s$^{-1}$ kpc$^{-1}$ and $B\approx-12$ km s$^{-1}$ kpc$^{-1}$;][]{Feast1997,Bovy2017, Donlon2024}.

\subsection{Radial acceleration}

\begin{figure}
    \centering
    \includegraphics[width=\columnwidth]{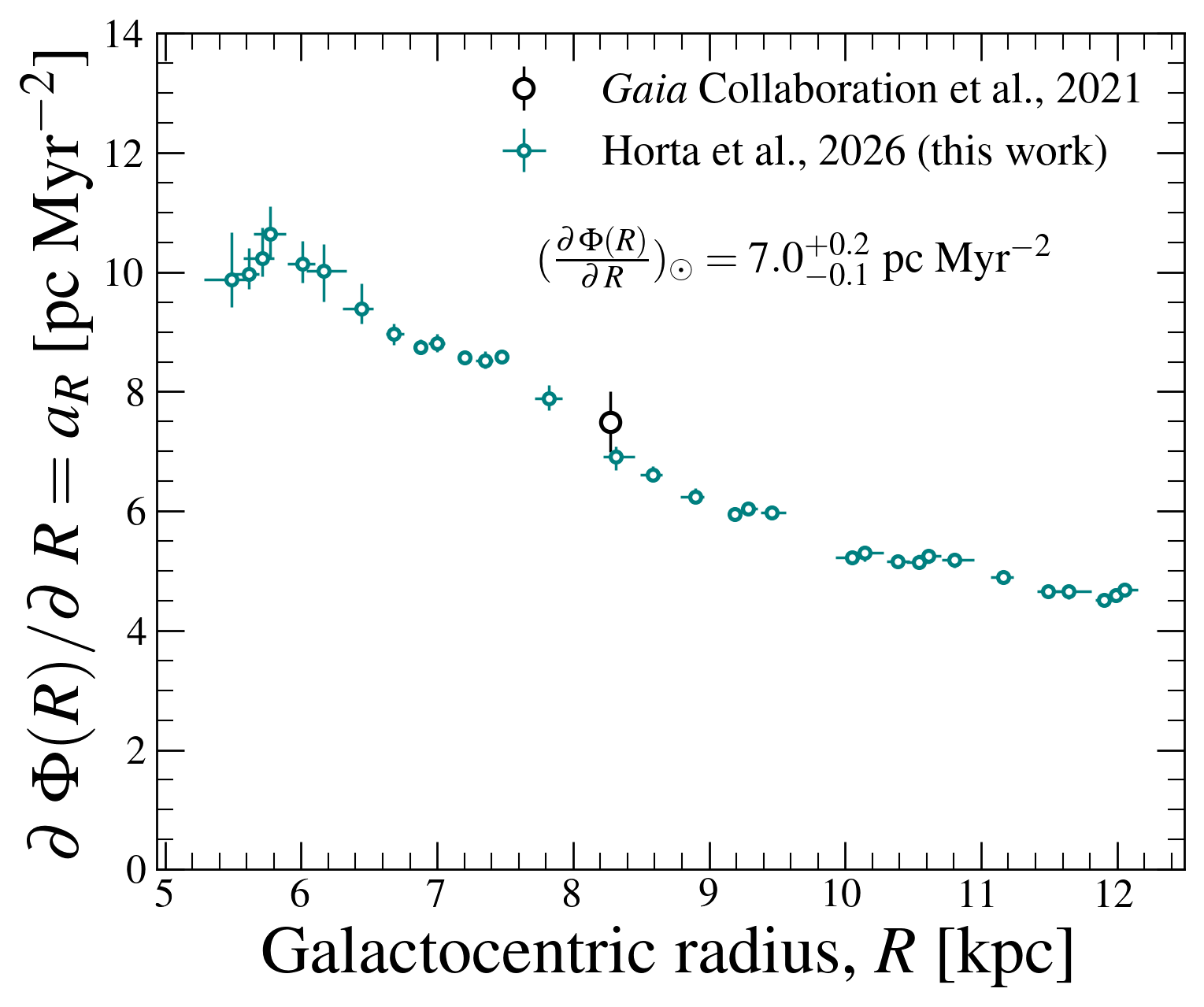}
    \caption{Measurements of the radial acceleration at every radius computed using Eq~\ref{eq_radacc}. We find that the radial acceleration drops from a value of $\partial\Phi/\partial R \approx10$ pc Myr$^{-2}$ at $R\approx6$ kpc to a value of $\partial\Phi/\partial R \approx 4.5$ pc Myr$^{-2}$ at $R\approx12$. At the Solar radius, we determine a value of $(\partial\Phi/\partial R)_{\odot} =7.0^{+0.2}_{-0.1}$ pc Myr$^{-2}$, which is in agreement with previous measurements \citep[][]{Gaia2021_acc}. }
 
    \label{fig_dphidR}
\end{figure}

 Figure~\ref{fig_dphidR} shows our measurement of the radial acceleration, $\partial\Phi/\partial R = a_R$, at every radius computed using the following relation
\begin{equation}
\label{eq_radacc}
    \frac{\partial \Phi}{\partial R}\bigg\rvert_{L_z} = \frac{v_c^2}{R} \bigg\rvert_{L_z}.
\end{equation}

This relation holds for an axisymmetric gravitational potential of a disk galaxy, evaluated close to the Galactic plane in a narrow slice of $L_z$. Given our results, we find that the radial acceleration drops from a value of $\partial\Phi/\partial R \approx10$ pc Myr$^{-2}$ at $R\approx6$ kpc to a value of $\partial\Phi/\partial R \approx 4.5$ pc Myr$^{-2}$ at $R\approx12$. At the Solar radius, we determine a value of $(\partial\Phi/\partial R)_{\odot} =7.0^{+0.2}_{-0.1}$ pc Myr$^{-2}$, which is in agreement with previous measurements \citep[e.g., $7.5\pm0.52$ pc Myr$^{-2}$;][]{Gaia2021_acc}.

\begin{table*}
	\centering
	\caption{From left to right, estimates of the circular velocity curve, radial acceleration, radial velocity, epicyclic/azimuthal frequencies, and Oort constants at the Solar radius estimated in this work using element abundance gradients.}
	\begin{tabular}{lcccccr} 
		\hline
    $v_{c,\odot}$ [km s$^{-1}$] & $a_{R_\odot}$ [pc Myr$^{-2}$]& $v_{R_{\odot},0}$ [km s$^{-1}$]& $\kappa_{0,R_\odot}$ [km s$^{-1}$ kpc$^{-1}$]& $\Omega_{0,R_\odot}$ [km s$^{-1}$ kpc$^{-1}$]& $A$ [km s$^{-1}$ kpc$^{-1}$]& $B$ [km s$^{-1}$ kpc$^{-1}$]\\
   \hline
   $235.3^{+2.8}_{-3.7}$ & $7.0^{+0.2}_{-0.1}$ & $-2.6^{+0.6}_{-1.1}$& $36.9^{+0.8}_{-1.0}$ & $28.5_{-0.1}^{+0.4} $& $16.5^{+0.1}_{-0.1}$ & $-11.9^{+0.1}_{-0.3}$\\
     \hline
	\end{tabular}
 \label{tab:results}
\end{table*}

\section{Discussion}
\label{sec_discussion}

\subsection{The Milky Way disk in chemical-kinematic space}
Figure~\ref{fig_grad_R_z} through Figure~\ref{fig_abun_lz_bins} illustrate that the Milky Way's low-$\alpha$ disk is well ordered in terms of its chemical and kinematic properties. Specifically, $\langle\mathrm{[Fe/H]}\rangle$ correlates with Galactocentric radius or angular momentum in the Galaxy, whereas $\langle\mathrm{[Mg/Fe]}\rangle$ correlates with radial (and vertical, \citealt{Horta2024_oti}) excitation. These trends are the same as the previously observed radial gradient in mean [Fe/H] with present day radius \citep{Cheng2012, Frankel2018, Eilers2022} and the inverse gradient in mean [Mg/Fe] with vertical position \citep{Schlesinger2014,Hayden2015, Bovy2016} in the Milky Way low-$\alpha$ disk. 
These chemical--kinematic trends could be a direct consequence of the inside-out and upside-down formation of galaxy disks \citep{Bird_2013}, modulo heating and migration processes.

Radial migration effects, either from ``blurring'' (radial heating leading to an increase in $J_R$) or ``churning'' (changes in angular momentum, $L_z$), could lead to chemical gradients being altered or washed out. Of course, our initial selection of stars with low radial and vertical action naturally leads to a sample that has preferentially suffered less from processes that drive blurring. While the primary dynamics responsible for churning \citep{Sellwood2002} is significantly more effective in populations with low radial action \citep{Daniel2018}, any scenario that drives disk wide churning would also cause significant kinematic heating. However, for old stars like the majority of the red giants modeled in this study, the fact that we observe smooth chemical abundance gradients in the plane of radial kinematics, and across the Galaxy more broadly, must place some limit on the magnitude of radial migration in our low-$\alpha$ disk sample, as such gradients will get washed out as radial migration proceeds \citep{Frankel2018,Frankel2020}.

The precision of this method depends on the steepness of the element abundance gradient with radial action.
It is possible that, for future applications with larger samples of stars, it would help this framework to select stars with younger ages that may have suffered from less blurring and thus may have steeper abundance gradients.

\subsection{A fully empirical estimate of the Milky Way circular velocity curve using element abundances}

The results presented in Figure~\ref{fig_vc} present a measurement of the circular velocity curve of the Milky Way determined using stellar chemical abundance-kinematic data. Circular velocity curve studies of the Milky Way date back to the early 20$^{\mathrm{th}}$ century \citep{Kapteyn1911, Oort1927a, Oort1927b}. Since then, measurements have been determined using either stars as tracers \citep{Bovy2012,Huang2016, Eilers2019, Mroz2019, Poder2023,Zhou2023,Ou2024,Fedorov2025}, gas \citep{Blitz1979,Fich1989,Merrifield1992, Honma1997, Sofue2009, Hou2014}, or both simultaneously \citep{Bhattacharjee2014, Reid2019}.
These prior methods require a model or correction for ``asymmetric drift,'' the effect that any population of stars with finite non-circular excitation will not orbit (on average) at the circular velocity. The method presented here requires no such correction, as the circular orbit is directly indicated in the abundance maps. While still restricted by the assumptions described in Section~\ref{sec_assumptions}, data-driven methods like the OTI framework presented in this study are advantageous as they provide a pathway to measure the structure of the Milky Way using data-driven approaches, exploiting this crucial relationship between the chemistry and kinematics of low-$\alpha$ disk stars, without the need to account for complex spatial selection function effects. So long as the abundance gradient is captured by the model in kinematic space and abundances do not depend on phase-space location, it is possible to empirically extract the orbital structure and mass density distribution (see \citealt{Horta2024_oti} and \citealt{Price2025} for further details). Moreover, OTI requires modelling a large ensemble of stars, thus its solutions are statistically meaningful. 

Our measurements of the circular velocity curve (Figure~\ref{fig_vc}) are in good agreement with the values measured in \citet{Eilers2019} for $8 < R < 12$ kpc. However, they appear to be in slight disagreement with the estimates towards the inner regions of the Milky Way disk. The likely reason for this is that, in contrast to other measurements that model the moments of kinematic tracers and produce an overall profile, our measurements of the circular velocity curve are determined using data in small bins of angular momentum. They are also computed empirically using a data-driven method that models the inflection point of the mean chemical abundance distribution w.r.t. kinematic observables. Compared to previous methods, our model is more flexible as it is potential independent and models directly observable data, enabling a measurement of the circular velocity curve with fewer assumptions. However, our measurement of the circular velocity curve also relies on the ability of the OTI model to identify the saddle point in the chemical abundance distribution for stars in a given $L_z$ bin in $R$ vs $v_R$ space (Figure~\ref{fig_contours}), and our assumption that this saddle point in the abundances corresponds to the point in the orbital distribution where $J_R = 0$ and $R=R_g$. Given that we are using near-infrared observations from the \textsl{APOGEE} survey, which are less affected by dust extinction than other wavelengths and can thus probe the Galactic disk much easier, we argue that we have the required data to robustly fit a model to a high degree of precision; this is supported by our results from a test on a smooth Milky Way model simulation (see Appendix~\ref{app_simulations_smooth}) or a live hydrodynamical galaxy simulation (see Appendix~\ref{app_simulations_live}). Thus, the discrepancies between the \citet{Eilers2019} circular velocity curve and our measurements are not likely due to biases in our model or data.

Intriguingly, in the inner disk, we observe some ``jumps'' in the values of $v_c$ determined by our method, which then results in our estimates being higher than those provided by \citet{Eilers2019}. We hypothesise that the offset between our model values in the inner disk is likely due to our model being able to capture dynamical perturbations, which traditional recipes like Jeans modeling are unable to account for. To test this hypothesis, in Appendix~\ref{app_simulations_live} we tested our model's ability to recover the circular velocity curve using a live hydrodynamic galaxy simulation from Hunt et al,. (in prep). The chemical and kinematic properties of this simulated galaxy qualitatively resemble the Milky Way at present day. Overall, our method seems to capture the circular velocity curve well, even if the system is in disequilibrium (Figure~\ref{fig_vc_sim_perturbed}). This has also shown to be the case for the plane of vertical kinematics when tested on live cosmological simulations of Milky Way-mass galaxies \citep{Oeur2025}. Interestingly, our results show that our method also predicts jumps in the measurements of $v_c$ in this live simulation, and these jumps align with the radius at which there is an overdensity of stars in that simulation (i.e., a spiral arm or dynamical resonance, around $R\approx9.2$ kpc). This result could explain why we see these jumps in Figure~\ref{fig_vc} for Milky Way data, and could be the result of our OTI method capturing overdensities or dynamical disequilibrium features in the Galactic disk (for example, the Scutum spiral arm; \citealp{Reid2019,Castro2021}). However, we note that the discrepancies between previous results and ours are small ($<10\%$), with maximal differences on the order of $\approx20$ km s$^{-1}$ in any given $L_z$ bin. Thus, such dynamical disequilibrium features must have a small effect.

All our measurements for stars at the Solar radius are summarized in Table~\ref{tab:results}.

\subsection{The impact of disequilibrium and non-axisymmetric features}
\label{sec_discussion_diseq}

The effect of disequilibrium and non-axisymmetric features in the Milky Way disk, arising either from the influence of the Milky Way bar, spiral arms, previous interactions with satellite galaxies and the aftermath of galaxy mergers, or all/some simultaneously, could affect the measurements of the circular velocity curve, epicyclic and azimuthal frequencies, and the kinematic gradients when modelled under certain assumptions of axisymmetry \citep[][]{Khoperskov2022,Almannaei2024}. When testing our method on a live hydrodynamic galaxy simulation, we have found that dynamical disequilibrium features (e.g., spiral arms) can induce small jumps in the $v_c$ value measured (Figure~\ref{fig_vc_sim_perturbed}), highlighting that our method captures these features. This result implies that our method, which uses unbiased stellar label kinematic tracers in small radial bins across the disk, is able to capture possible deviations from axisymmetry. This is because our measurement provides an inference of the \textit{instantaneous} value of these dynamical quantities, instead of an averaged or analytical value, as typically derived using traditional methods (such as the gray lines in Figure~\ref{fig_vc}, Figure~\ref{fig_vc_sim}, or Figure~\ref{fig_vc_sim_perturbed}). This results in our method being able to identify regions in the disk where these assumptions fail. However, given that we are working under the assumption of equilibrium and steady-state, our model is unable to explain these deviations.

\subsection{Future prospects and limitations}

In this work we have set out to model the plane of radial kinematics in the Milky Way's low-$\alpha$ disk using one element abundance ratio, [Mg/Fe], that given our tests provides the strongest gradient (see Figure~\ref{fig_abun_all_relations} in the Appendix). However, large spectroscopic surveys (like \textsl{APOGEE}) provide detailed element abundance ratios for a spectrum of elements, which should also display gradients. In principle, all element abundance ratios should provide constraints on the orbital shapes and mass distribution. However, many element abundance species are produced in similar nucleosynthetic channels and are thus not independent \citep[e.g.,][]{Ness2022}. In future work, it will be interesting to identify and use independent combinations of element abundance ratios as our stellar labels that track different nucleosynthetic channels (e.g., SNe II, SNe Ia, AGB winds, etc...), following a similar approach to the KPM model \citep{Griffith2024}. Synergising large spectroscopic surveys that are either currently active (\textit{APOGEE}: \citealp{Majewski2017}; \textit{Gaia}: \citealp{Gaia2022}; \textit{GALAH}: \citealp{Buder2022}; \textit{LAMOST}: \citealp{Cui2012}; \textit{DESI}: \citep{Cooper2023}) or that will become active in the coming years (e.g., \textit{Milky Way Mapper}: \citealp{Kollmeier2017}; \textit{WEAVE}: \citep{Dalton2012}; \textit{4MOST}: \citealp{DeJong2012}) with statistical and machine-learning methods \citep[e.g.,][]{Ness2015,Ting2019, Leung2024, Horta2025_lux} will provide a path towards obtaining high-precision measurements across a wide range of elements, perfect for performing a joint inference. Of particular importance will be the \textit{Milky Way Mapper} survey, that will imminently deliver \textit{APOGEE}-resolution full-sky spectra for $>3$ million Milky Way red giant stars; similarly, the vast amount of medium-resolution RVS spectra that \textit{Gaia} DR4 will deliver will be extremely useful for OTI and methods like the one presented in this study. Building on the work from \citet{Horta2024_oti}, in the future it will be useful to use these data to not only measure the inflection point in the mean abundance trend with radial kinematics, but to create a radial OTI model for the Milky Way disk in order to capture the full distribution of orbits from element abundance measurements. Such an endevour would provide an additional way of performing the measurements made in this study by capturing the morphology and foliation of radial orbits described by element abundance gradients. 

Insofar, we have limited our analysis to examining a two-dimensional kinematic plane (in this work, the radial one). However, building an OTI model to characterise fully the plane of radial kinematics, and combining it with the vertical kinematic case described in \citet{Price2025} and tested \citet{Horta2024_oti}, would provide an avenue to develop an OTI framework that is capable of modelling the 4D phase-space directly. This would enable the model to be free of the assumption of separability of the potential in the radial and kinematic directions for Milky Way disk stars (i.e., $\Phi (R,z) = \Phi(R) + \Phi(z)$). For stars on near-circular orbits in the Milky Way low-$\alpha$ disk, one way this could be achieved is by including higher order and cross terms when expanding in Taylor series the effective potential, such that
\begin{align}
    \Phi_{\mathrm{eff}} = \Phi_{\mathrm{eff}}(R_g,0) + \frac{1}{2}\Big(\frac{\partial^2\Phi_{\mathrm{eff}}}{\partial R^2}\Big)_{(R_g,0)}x^2 + \\\frac{1}{2}\Big(\frac{\partial^2\Phi_{\mathrm{eff}}}{\partial z^2}\Big)_{(R_g,0)}z^2 + O(xz^2).
\end{align}
The inclusion of the $O(xz^2)$ term would allow one to capture the curvature in the orbital shape of stars on non-perfectly circular orbits (i.e., the curve in the trapezoid delineated by a non-boxy orbit). However, disentangling how to interpret this to make robust measurements of the enclosed mass would become a challenge. We reserve this for future exploration.

\section{Conclusions}
\label{sec_conclusions}
In this paper we use \textsl{APOGEE} and \textsl{Gaia} data to model the plane of radial kinematics and measure the Milky Way's circular velocity curve. To do so, we have modelled the inflection point of the mean element abundance trend in the plane of radial kinematics (i.e., $R$-$v_R$), using a tweaked version of the Orbital Torus Imaging modeling framework (OTI; \citealp{Price2021, Price2025}). Instead of resorting to a parameterized model of the global Galactic potential, we partition the data into bins of angular momentum to perform our measurements with the radial phase-space (kinematic) coordinates across small regions of the disk. Our main results and findings are summarized below.

\begin{enumerate}
    \item For a narrow bin in angular momentum, the plane of radial kinematics in the low-$\alpha$ disk can be modeled using the average [Mg/Fe] abundance of stars, as this element abundance ratio traces the radial excursions around a guiding-center radius (Figure~\ref{fig_contours}). This explains vertical gradient in $\langle$[Mg/Fe]$\rangle$ in the Milky Way low-$\alpha$ disk (Figure~\ref{fig_grad_R_z} through Figure~\ref{fig_abun_lz_bins}). Conversely, across the Galactic low-$\alpha$ disk, the radial (or angular momentum) dependence can be traced with $\langle$[Fe/H]$\rangle$, which leads to the known metallicity-radius gradient.
    \item We are able to empirically measure the Milky Way circular velocity curve across the disk using stellar labels (Figure~\ref{fig_vc}). Our results for the circular velocity curve closely match that of previous works for stars at the Solar radius ($v_{c,\odot} = 235.3^{+2.8}_{-3.7}$ km s$^{-1}$) and beyond, but show small discrepancies ($<10\%$) for the inner regions of the disk. We have validated that our method works by testing it on a controlled simulation of a smooth Milky Way-mass model (Figure~\ref{fig_vc_sim} in Appendix~\ref{app_simulations_smooth}) and on a live hydrodynamic galaxy simulation (Figure~\ref{fig_vc_sim_perturbed} in Appendix~\ref{app_simulations_live}).
    \item Given our measurement of the circular velocity curve, we are also able to empirically measure the radial and azimuthal frequencies across the Milky Way low-$\alpha$ disk, and by default kinematic gradients (i.e., Oort constants, Figure~\ref{fig_freq}). Overall, we find good agreement between our measurements and that of previous work, yielding values at the Solar radius of: $\kappa_{0,R_\odot}= 36.9^{+0.8}_{-1.0}$ km s$^{-1}$ kpc$^{-1}$; $\Omega_{0,R_\odot}= 28.5_{-0.1}^{+0.4}$ km s$^{-1}$ kpc$^{-1}$; $A = 16.5^{+0.1}_{-0.1}$; $B = -11.9^{+0.1}_{-0.3}$. 
    \item We are also able to determine the radial acceleration in each bin of angular momentum (Figure~\ref{fig_dphidR}). At the Solar radius, our measurement of $(\frac{\partial \Phi}{\partial R})_{\odot} = a_{R_\odot} = 7.0^{+0.2}_{-0.1}$ pc Myr$^{-2}$ is in agreement with previous measurements.
\end{enumerate}

\section*{Acknowledgements}
DH would like to thank Larry Widrow, Kathryn Johnston, the Dynamics tea group at the Royal Observatory of Edinburgh, and the Nearby Universe group at the Center for Computational Astrophysics at the Flatiron Institute for helpful discussions. He would also like to thank Sue, Alex, and Debra for all their support. 
This work was supported by the UKRI Science and Technology Facilities Council under project 101148371 as a Marie Sk\l{}odowska-Curie Research Fellowship. AMPW and DWH note that the Flatiron Institute is a division of the Simons Foundation. SEK acknowledges support from the Science $\&$ Technology Facilities Council (STFC) grant ST/Y001001/1. JH acknowledges the support of a UKRI Ernest Rutherford Fellowship ST/Z510245/1. DWH thanks Neige Frankel for valuable conversations. KJD acknowledges support from the Heising Simons Foundation grant \# 2022-3927. AMPW and DWH respectfully acknowledge that the land politically designated as New York City is the homeland of the Lenape (Lenapehoking), who were unjustly displaced. KJD respectfully acknowledges that the University of Arizona is home to the O'odham and the Yaqui. She respects and honors the ancestral caretakers of the land, from time immemorial until now, and into the future.

\section*{Data Availability}
All \textsl{APOGEE} and \textsl{Gaia} data used in this study are publicly available and can be downloaded directly from \url{https://www.sdss4.org/dr17/} and \url{https://gea.esac.esa.int/archive/}, respectively.

\software{
    \texttt{AGAMA}: \citep{Vasiliev2019},
    \texttt{Astropy}: \citep{astropy2013, astropy2018, astropy2022},
    \texttt{gala}: \citep{gala},
    \texttt{JAX}: \citep{jax2018github},
    \texttt{JAXOpt}: \citep{jaxopt:2021},
    \texttt{matplotlib}: \citep{Hunter:2007},
    \texttt{numpy}: \citep{numpy},
    \texttt{pyia}: \citep{pyia},
    \texttt{scipy}: \citep{scipy}.
}

\bibliography{refs}

\appendix

\section{$\langle$[Fe/H]$\rangle$ abundance gradients in radial kinematics}
\label{app_lzbins}

\begin{figure*}
    \centering
    \includegraphics[width=\textwidth]{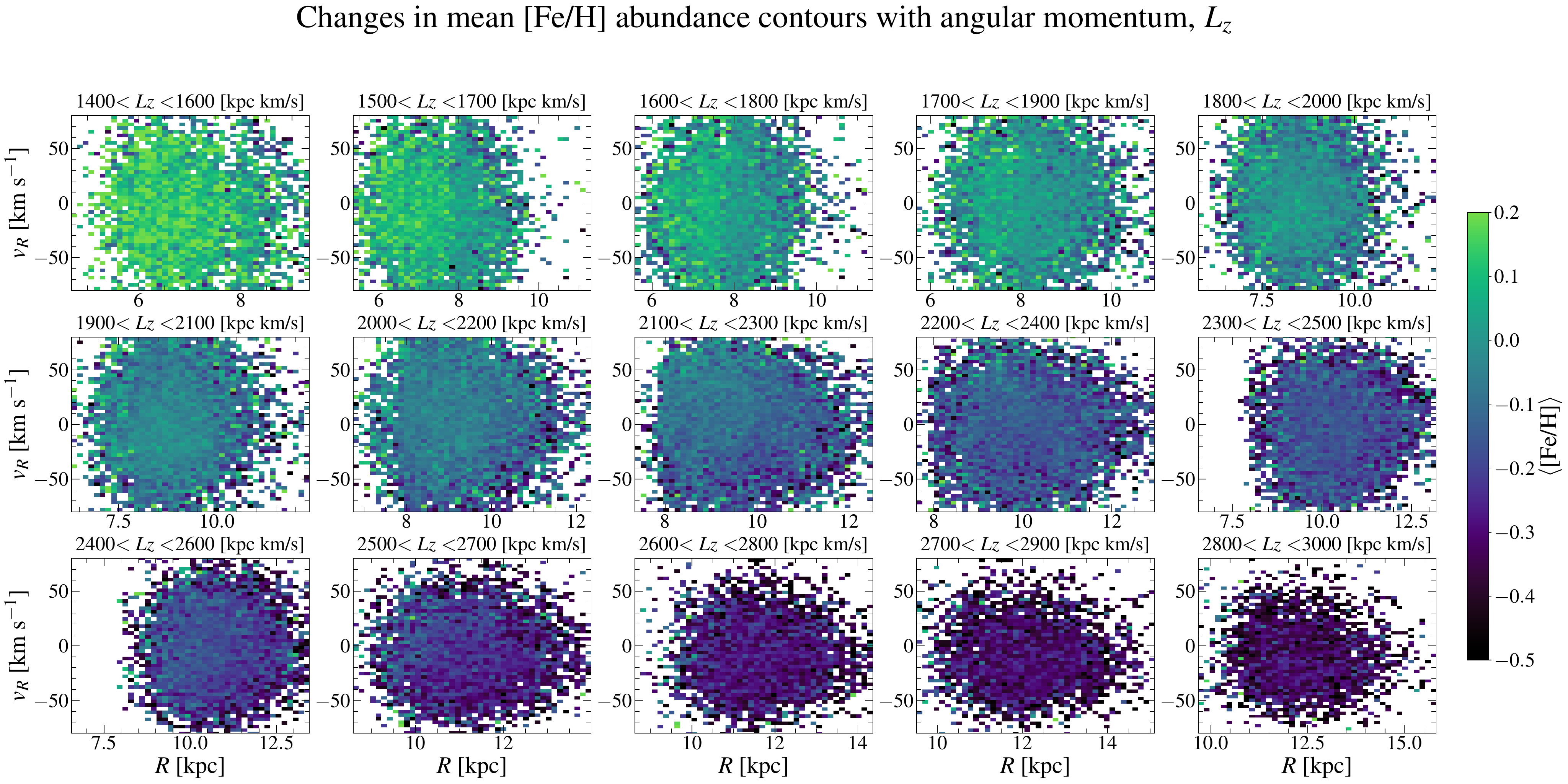}
    \caption{Same as Figure~\ref{fig_abun_lz_bins} but now showing the average [Fe/H] value in each pixel. $\langle$[Fe/H]$\rangle$ does not display any mono-abundance gradients in bins of $L_z$. Conversely, in each bin, the value of $\langle$[Fe/H]$\rangle$ is approximately equal for all values of $R$ and $v_R$. However, there is a $\langle$[Fe/H]$\rangle$ gradient with $L_z$, which is to be expected given the radius-[Fe/H] gradient in the low-$\alpha$ disk \citep[see Figure~\ref{fig_R_vphi} and ][]{Frankel2019}.}
    \label{fig_app_lz_feh}
\end{figure*}

\clearpage

\section{Choosing the best element abundance ratio}
\label{app_elements}

\begin{figure*}
    \centering
    \includegraphics[width=\textwidth]{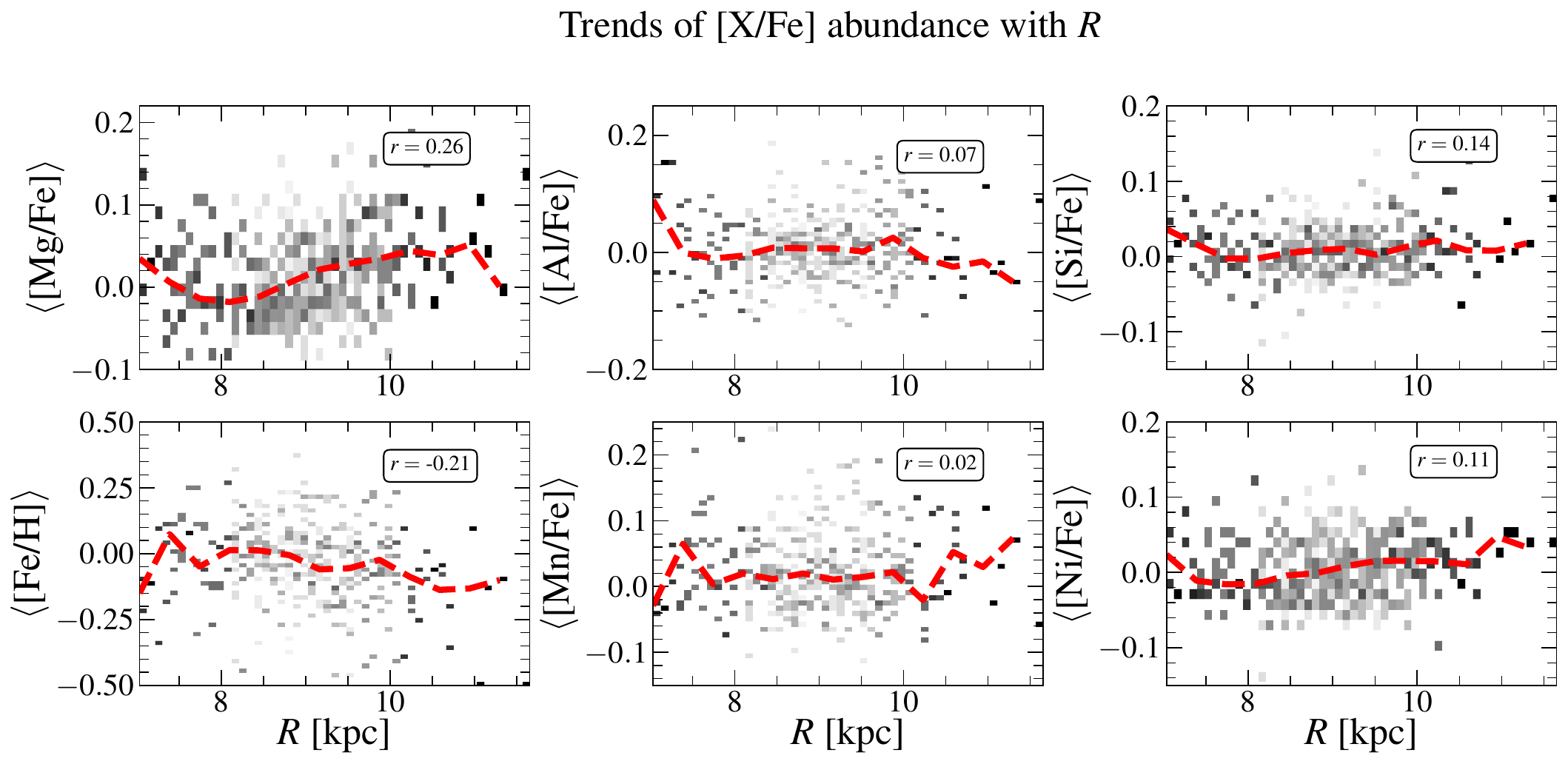}
    \caption{Test to identify the element abundance ratio that shows the best correlation with radial kinematics. Here, the 2D density distribution is shown for all stars at the Solar radius from our working sample with $1,900 < L_z < 2,050$ kpc km s$^{-1}$, $|v_R|<10$ km~s$^{-1}$, $|v_z|<10$ km~s$^{-1}$, and $|z|<0.1$ kpc. To quantify the level of correlation, for each of the six abundance probed we compute the Pearson correlation function value, $r$. We only chose to study six of the most well measured elements in \textsl{APOGEE}. Similarly to when modeling the plane of vertical kinematics \citep{Horta2024_oti}, we find that $\langle$[Mg/Fe]$\rangle$ shows the strongest correlation with radial kinematics for low-$\alpha$ disk stars. Thus, we choose to use this abundance ratio in our modelling. }
        \label{fig_abun_all_relations}
\end{figure*}

Figure~\ref{fig_abun_all_relations} shows a 2D density distribution of Milky Way low-$\alpha$ disk stars modelled. Specifically, each panel shows the correlation between average abundance as a function of $R$, which for a narrow slice in $|v_R|$ around $v_R\approx0$, this is analogous to the maximum value of the epicyclic excursion (i.e., difference between present day radius, $R$, and guiding-center radius, $R_g$, so $R\sim \Delta R_{\mathrm{max}}$). We only tested six of the best elements determined in the \textsl{APOGEE} survey (namely, [Mg/Fe], [Al/Fe], [Si/Fe], [Fe/H], [Mn/Fe], and [Ni/Fe]). As a dashed red line we show the running median. To quantify the level of correlation to pick the best element abundance ratio, we compute the Pearson correlation function value, $r$. Our results indicate that average [Mg/Fe] is the element with the strongest correlation with radial kinematics. 

\clearpage

\section{Tests on controlled simulations}
\label{app_simulations}

\subsection{Tests on a smooth axisymmetric Milky Way model}
\label{app_simulations_smooth}

In this section we go on to test how well we are able to recover these quantities in a controlled simulation. To do so, we use the axisymmetric quasi-isothermal disk distribution function \citep[DF:][]{Binney2012} from \citet{Price2025}, that employs parameters derived from fitting Milky Way stellar data. This DF is implemented using the \texttt{Agama} package \citep{Vasiliev2019}, sampling $5\times 10^8$ phase-space coordinates with guiding-center radii, $R_g$, between $5<R_g<12$ kpc, embedded in a simple Milky Way potential model. This model uses a two-component set-up consisting of a Miyamoto-Nagai disk \citep{Miyamoto1975} embedded in a spherical NFW halo \citep{Navarro1997}; the model adopts the following parameter values for the disk and halo, respectively: $M_{\mathrm{disk}} = 6.91 \times 10^{10} M_{\odot}$, $a=3$ kpc (scale length of the disk), $b=0.25$ kpc (scale height of the disk), $M_{\mathrm{halo}}= 5.4 \times 10^{11} M_{\odot}$ (scale mass of the halo), and $r_s = 15$ kpc (scale radius of the halo). These values were chosen by \citet{Price2025} so that the circular velocity at the solar radius, $R_{0} = 8.3$ kpc is $v_c(R_0) = 229$ km s$^{-1}$, and other parameters are consistent with the \texttt{MilkyWayPotential2022} implemented in \texttt{gala} \citep{Price2017}.

\subsubsection{OTI model set-up}
In order to resemble the selection for the low-$\alpha$ disk employed in the \textsl{APOGEE-Gaia} data, we enforce the following cuts on our simulation data sample: $|z|<0.5$ kpc, $|v_z|<30$ km~s$^{-1}$, $|v_R|<100$ km~s$^{-1}$. These selections enable us to restrict our sample to stars with lower radial action that are more confined to the midplane, thus bringing us closer to our assumption of $R$-$z$ separability. These selection criteria yield $\approx5,000,000$ star particles.

We compute simulated [Mg/Fe] abundance values for these star particles using the linear relation between [Mg/Fe] and radial action, $J_R$, from the \textsl{APOGEE} data, including a scatter of $0.05$ dex, with simulated uncertainties following the prescription from \citet{Price2025}. 

We then model the simulated mean abundance data in pixels of phase-space coordinates using the framework described in Section~\ref{sec_model}. To model the uncertainties, we bootstrap sample with replacement the phase-space and chemical abundance values (namely, $\{R,v_R,\mathrm{[Mg/Fe]},\sigma_{\mathrm{[Mg/Fe]}}\}$) for $25$ distinct samples of $5\times 10^4$ star particles across 32 different angular momentum bins, spanning from $1,200$--$2,750$ kpc~km~s$^{-1}$ (each of 100 kpc~km~s$^{-1}$ width and separated by 50 kpc~km~s$^{-1}$). To provide flexibility in the model, we use monotonic quadratic spline functions for both the label function and for the Fourier coefficient functions. For the label function, we use eight knots equally spaced in $r_R$ between $0$ and $r_{R,\mathrm{max}}$. Conversely, for the Fourier terms $m=\{1,2,3,4\}$ we use $\{8,12,4,4\}$ knots, respectively. We space the Fourier coefficient spline knots equally in $r_R^2$ between $0$ and $r_{R,\mathrm{max}}$; this places a higher density of knots at lower $r_R$ values where we expect the Fourier coefficient functions to change more rapidly. We again require that the coefficient function values are zero at $\tilde{r_R}=0$ so that $e_1(0)= e_2(0)=e_3(0)=e_4(0)=0$. We assume that the overall sign of the functions are ($-,+,-,+$) for the $m=\{1,2,3,4\}$ coefficient functions, respectively.

To optimize, we follow the method described in \citet{Price2025} and use a Gaussian log-likelihood for the mean [Mg/Fe] abundance data in each pixel of phase-space. Thus, the log-likelihood of the data, $\ln \mathcal{L}$, is computed as
\begin{equation}
    \ln \mathcal{L} = \sum_j \ln \mathcal{N}(\langle Y_{\mathrm{[Mg/Fe]}}\rangle_j|Y_j,\sigma_{Y,j}),
\end{equation}
where the sum is done over the $j$ pixels. $\langle Y_{\mathrm{[Mg/Fe]}}\rangle_j$ is the mean abundance value in pixel $j$, $Y_j$ is the model predicted mean abundance value in pixel $j$, and $\sigma_{Y,j}$ is the uncertainty on the mean abundance. For the uncertainty on the mean abundance in each pixel we consider both the error on the mean
\begin{equation}
\sigma_{\mu_Y} = \sqrt{\frac{1}{\sum_{i}\frac{1}{\sigma^2_{Y,i}}}},    
\end{equation}
and the intrinsic scatter of abundances in each pixel (see \citet{Price2025} for further details). We use a standard L-BFGS-B optimizer \citep{Byrd1995} to minimize the regularised negative log-probability as implemented in \texttt{JAXopt} \citep{jaxopt:2021}.

\subsubsection{Simulation results}

\begin{figure}
    \centering
    \includegraphics[width=0.5\columnwidth]{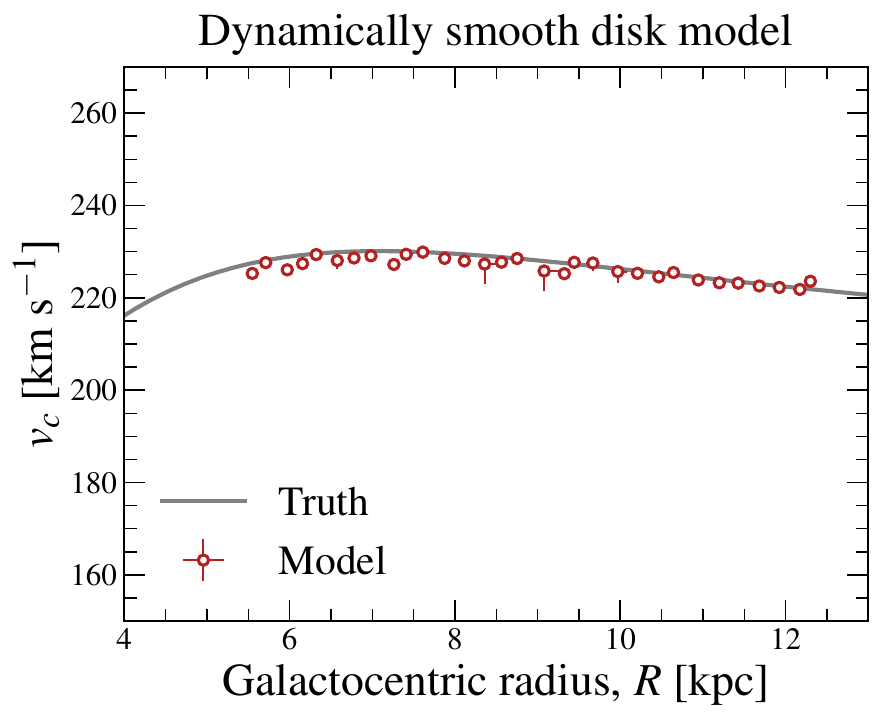}
    \caption{Measured circular velocity curve as a function of Galactocentric radius using our OTI model (Eq~\ref{eq_vc}) in red, contrasted with the true circular velocity curve in gray, for a test on simulated data in a live hydrodynamic galaxy simulation. The scatter points show the median and the error bars show the [16$^{\mathrm{th}}$, 84$^{\mathrm{th}}$] percentile uncertainty from the 25 sample bootstrap with replacement. The measurements of $v_c$ obtained with our framework are completely empirical, and are determined by finding the saddle point of the abundance gradient in $R$-$v_R$ space for stars in narrow intervals of angular momentum. Within the bootstrap uncertainties, our model recovers well the circular velocity curve measurement across a broad range of radii.}
    \label{fig_vc_sim}
\end{figure}

Figure~\ref{fig_vc_sim} shows the measured circular velocity curve using our method (Eq~\ref{eq_vc}) in red and the true circular velocity curve in black for a test on simulated data, assuming the circular velocity curve from \citep{Eilers2019}. The scatter points show the median and the error bars show the [16$^{\mathrm{th}}$, 84$^{\mathrm{th}}$] percentile uncertainty from the 25 sample bootstrap with replacement. The measurements of $v_c$ obtained with the OTI framework are completely empirical, and are determined by finding the saddle point of the abundance gradient in $R$-$v_R$ space for stars in narrow intervals of angular momentum. Within the bootstrap uncertainties, our model recovers well the circular velocity curve measurement across a broad range of radii. 

\subsection{Tests on a live hydrodynamic galaxy simulation}
\label{app_simulations_live}

We now go on to test how well we are able to recover these quantities in a live hydrodynamic galaxy simulation which should better represent the conditions in the Milky Way disc. To do so, we use the $\sim1\times10^8$ particle Milky Way-like host galaxy simulation from Hunt et al. (in prep), which consists of a live NFW dark halo \citep{Navarro1997}, a Miyamoto-Nagai stellar disc \citep{Miyamoto1975}, a stellar halo, a gas disk and a hot gas corona. The initial conditions were generated with \texttt{pgen}\footnote{https://github.com/kawatadaisuke/diskm-buildini} \citep{Kawata14}, and evolved for 8.1 Gyr using the hydrodynamical simulation code \texttt{ChaNGa}\ \citep{Jetley_Changa,Jetley_Changa2} with star formation, chemical evolution, and feedback enabled such that the disc contains self-consistently formed stellar populations. The galaxy develops a bar and spiral arms which cause radial migration and local radial disequilibrium features. A full analysis of this galaxy is beyond the scope of this work (see Hunt et al in prep for full details of setup and analysis), but provides a useful test for our method.

As in the previous Section, we make the following cuts to resemble our selection of the low-$\alpha$ disk employed in the \textit{APOGEE-Gaia} data: we selected stars that formed self-consistently within the simulation with $|z|<0.5$ kpc, $|v_z|<30$ km~s$^{-1}$, $|v_R|<100$ km~s$^{-1}$. This yields $\approx13,000,000$ star particles.

We repeat the bootstrapping procedure from the previous section, and determine 25 distinct samples of $5\times10^4$ star particles across the same 32 bins in angular momentum. However, instead of painting on the abundances as done in the smooth simulation, we instead use the [$\alpha$/Fe] values outputted from the simulation itself (tracked as [O/Fe] in \texttt{ChaNGa}), since this is a self-consistent hydrodynamic simulation. We enforce a value of 10$\%$ of the [$\alpha$/Fe] value as the chemical abundance uncertainty to mimic observational data. We then initialise the OTI model and optimise it in the same way, as described in the previous section.

\subsubsection{Simulation results}

\begin{figure}
    \centering
    \includegraphics[width=0.5\columnwidth]{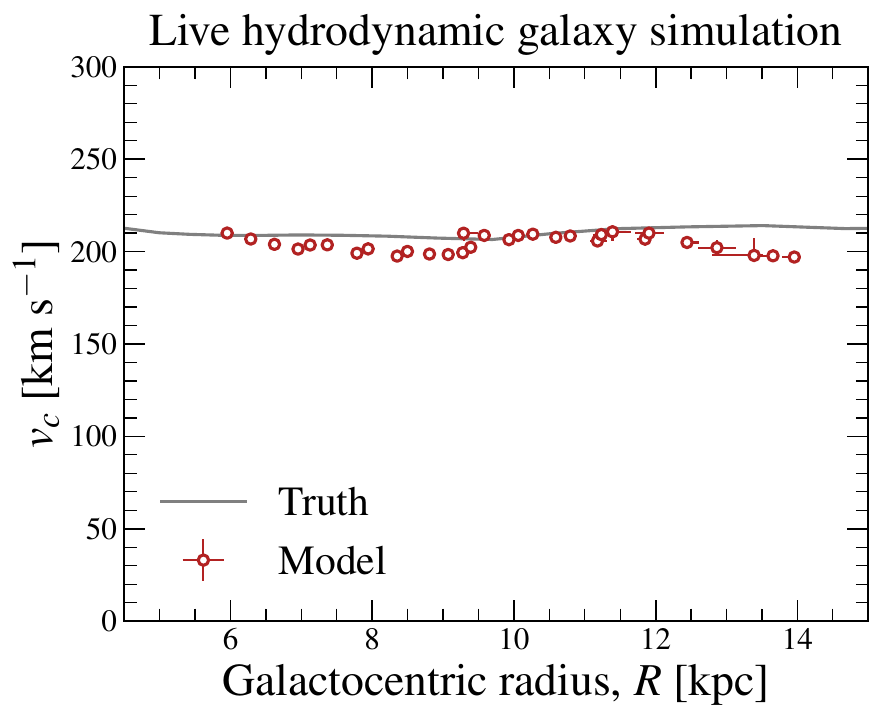}
    \caption{Measured circular velocity curve as a function of Galactocentric radius using our OTI model (Eq~\ref{eq_vc}) in red, contrasted with the true circular velocity curve in gray, for a test on simulated data in a live hydrodynamic galaxy simulation resembling the Milky Way system. The scatter points show the median and the error bars show the [16$^{\mathrm{th}}$, 84$^{\mathrm{th}}$] percentile uncertainty from the 25 sample bootstrap with replacement. The measurements of $v_c$ obtained with our framework are completely empirical, and are determined by finding the saddle point of the abundance gradient in $R$-$v_R$ space for stars in narrow intervals of angular momentum. Overall, our model recovers well the circular velocity curve measurement across a broad range of radii. However, we do note that due to dynamical disequilibrium features, there is a ``jump'' in our measurement of $v_c$ around $R\sim9.2$ kpc, that aligns with the radius at which there is an overdensity of stars (like that created by a spiral arm or a dynamical resonance). This feature could explain the jump we see in Figure~\ref{fig_vc} at $R\sim6$ kpc, that could be the result of the OTI model capturing a disequilibrium feature in the Milky Way low-$\alpha$ disk.}
    \label{fig_vc_sim_perturbed}
\end{figure}

\begin{figure}
    \centering
    \includegraphics[width=0.5\columnwidth]{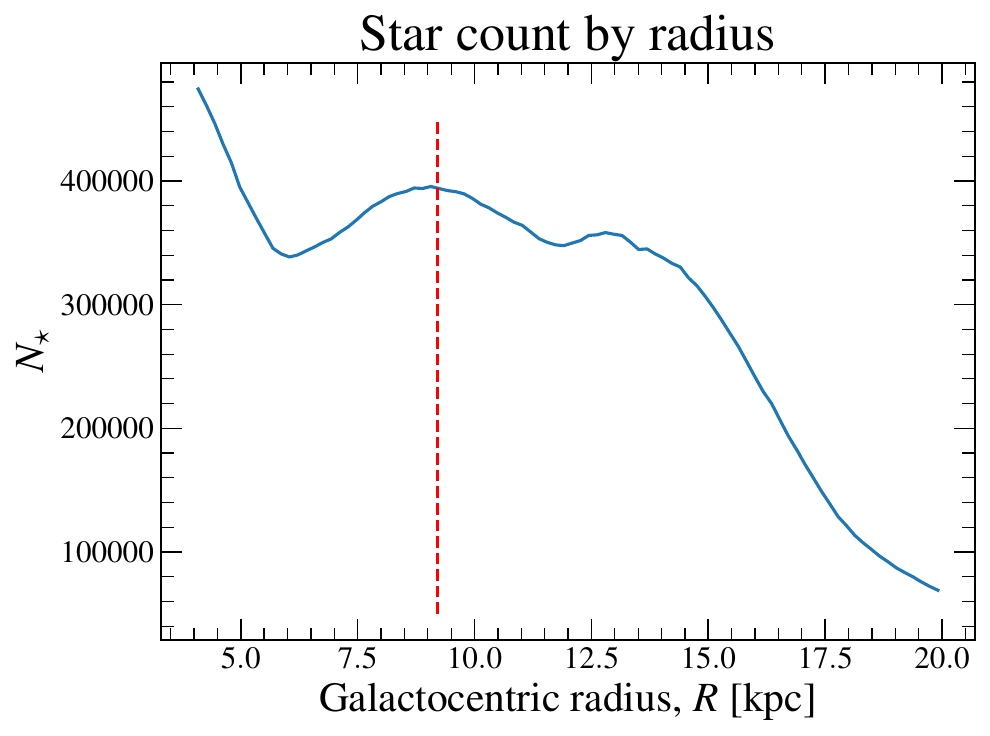}
    \caption{Number of star particles in the simulation in bins of Galactocentric radius. The vertical dashed red line highlights the radius at which we observe a jump in the inferred $v_c$ values from Figure~\ref{fig_vc_sim_perturbed}, that appears to coincide with an overdensity of stars.
    }
    \label{fig_app_sim_rgal}
\end{figure}

Figure~\ref{fig_vc_sim_perturbed} shows the measured circular velocity curve using our method (Eq~\ref{eq_vc}) in red and the true circular velocity curve in black for a test on simulated data. The scatter points show the median and the error bars show the [16$^{\mathrm{th}}$, 84$^{\mathrm{th}}$] percentile uncertainty from the 25 sample bootstrap with replacement. The measurements of $v_c$ obtained with the OTI framework are completely empirical, and are determined by finding the saddle point of the abundance gradient in $R$-$v_R$ space for stars in narrow intervals of angular momentum. Overall, our model recovers well the circular velocity curve measurement across a broad range of radii within the bootstrap uncertainties. However, we do note that due to dynamical disequilibrium features, there is a ``jump'' in our measurement of $v_c$ around $R\sim9.2$ kpc, that aligns with the radius at which there is an overdensity in the number of stars (like that created by a spiral arm or a dynamical resonance, see Figure~\ref{fig_app_sim_rgal}). This feature could explain the jump we see in Figure~\ref{fig_vc} at $R\sim6$ kpc, that could be the result of the OTI model capturing a disequilibrium feature in the Milky Way low-$\alpha$ disk.

While in this test we have found that an overdensity of stars could explain the jump in $v_c$, we are cautious about over-interpreting this result as a way of identifying spiral arms. This is because correlation doesn't necessarily imply causation, and other influences such as bar resonances could cause a similar effect. This is one potential explanation for what we see in the data, but it's not conclusive.

\end{document}

%% file: preamble.tex
\usepackage{savesym}
\savesymbol{tablenum}
\usepackage{siunitx}
\restoresymbol{SIX}{tablenum}
\DeclareSIUnit\year{yr}
\DeclareSIUnit\parsec{pc}
\DeclareSIUnit\msun{M_\odot}
\DeclareSIUnit\Rsun{R_\odot}

\newcommand{\kms}{\unit{\km\per\s}}

\DeclareSIUnit\mstar{M_\star}

\newcommand{\rR}{\ensuremath{r_R}}
\newcommand{\rRp}{\ensuremath{\tilde{r}_R}}

\newcommand{\thRp}{\ensuremath{\tilde{\theta}_R}}
